\documentclass[sigconf,nonacm]{acmart}

\AtBeginDocument{%
  }

\copyrightyear{2027}
\acmYear{2027}
\setcopyright{cc}
\setcctype{by}
\acmConference[CHI '27]{Proceedings of the 2027 CHI Conference on Human Factors in Computing Systems}{May 10--14, 2027}{Pittsburgh, PA}
\acmBooktitle{Proceedings of the 2027 CHI Conference on Human Factors in Computing Systems (CHI '27), May 10--14, 2027, Pittsburgh, PA}
\acmPrice{}
\acmDOI{XXXXXXX.XXXXXXX}
\acmISBN{978-1-4503-XXXX-X/2018/06}

\begin{document}

\title[Synthetic TLX]{Synthetic TLX: \\Forecasting Human Workload Using Agent Simulation} 


\author{Tzu-Sheng Kuo}
\authornote{Work done as a Student Researcher at Google DeepMind.}
\email{tzushenk@cs.cmu.edu}
\orcid{0000-0002-1504-7640}
\affiliation{
  \institution{Carnegie Mellon University}
  \city{Pittsburgh}
  \state{PA}
  \country{USA}
}

\author{Carrie J. Cai}
\email{cjcai@google.com}
\orcid{0000-0001-9421-7128}
\affiliation{
  \institution{Google DeepMind}
  \city{Mountain View}
  \state{CA}
  \country{USA}
}

\author{Meredith Ringel Morris}
\email{merrie@google.com}
\orcid{0000-0003-1436-9223}
\affiliation{
  \institution{Google DeepMind}
  \city{Seattle}
  \state{WA}
  \country{USA}
}

\author{Michael Terry}
\email{michaelterry@google.com}
\orcid{0000-0003-1941-939X}
\affiliation{
  \institution{Google DeepMind}
  \city{Cambridge}
  \state{MA}
  \country{USA}
}

\renewcommand{\shortauthors}{Kuo et al.}

\begin{abstract}
Assessing human workload for technology-mediated tasks helps prevent task failure caused by poor technology design. Traditionally, workload is assessed retrospectively using the NASA Task Load Index (TLX) after humans complete a task. What if we could forecast workload before a human attempts a task using agent simulation? We introduce Synthetic TLX, a new paradigm for proactive workload estimation that predicts NASA TLX scores for a given task, unlocking novel interaction opportunities and evaluation methods. To understand its viability, we conducted three experiments comparing human and agent-generated scores to evaluate where they align and diverge. We found agent estimates align with human scores particularly when prompted with a human persona and active task simulation. However, agents and humans diverge in the sources of workload they are sensitive to. Based on our findings, we present three applications to showcase Synthetic TLX's potential and discuss the future of workload-aware human-AI interaction.
\end{abstract}

\ccsdesc[500]{Human-centered computing~Human computer interaction (HCI)}
\ccsdesc[500]{Computing methodologies~Modeling and simulation}

\keywords{Synthetic TLX, NASA TLX, workload assessment, agent simulation}
\begin{teaserfigure}
  \vspace{2em} 
  \includegraphics[width=\textwidth]{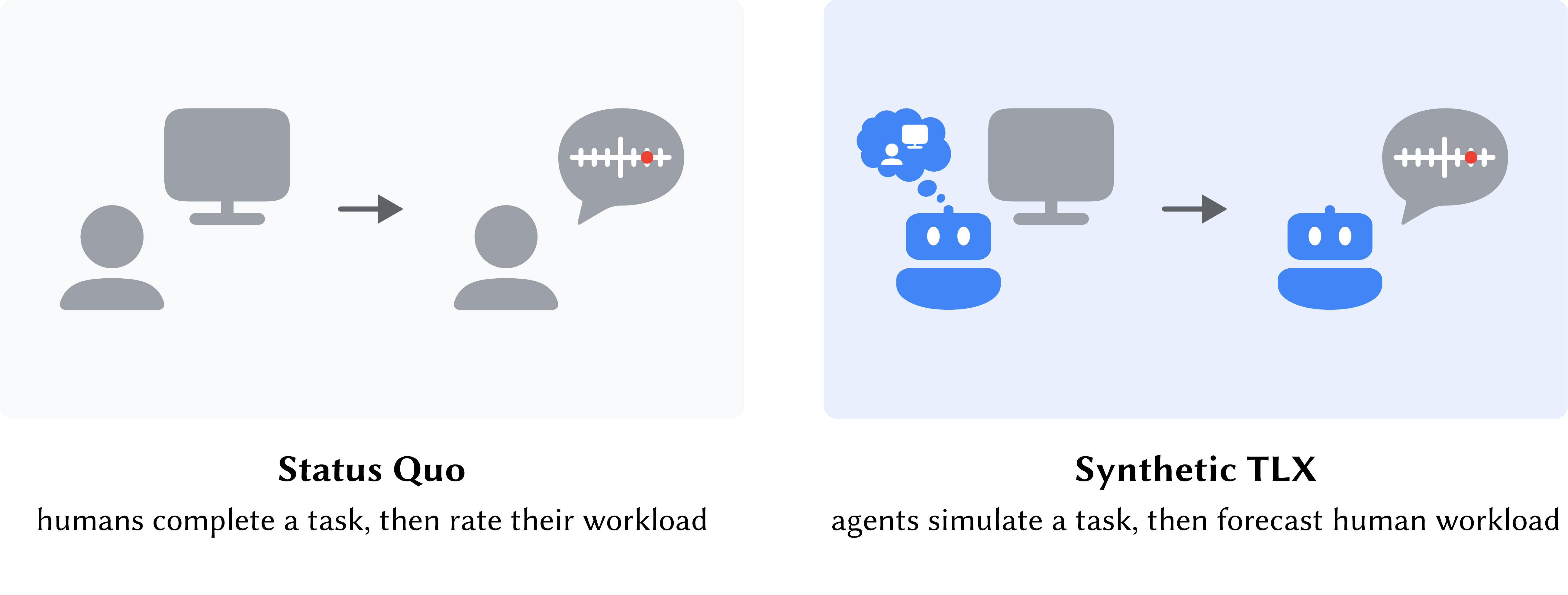} 
  \caption{We introduce Synthetic TLX, a proactive paradigm for human workload estimation. As illustrated, the status quo requires humans to complete a task and retrospectively rate their experienced workload. In contrast, Synthetic TLX uses AI agents to simulate a human executing the task and forecast the workload before a person even attempts it.}
  \Description{A side-by-side diagram comparing human workload assessment and agent workload estimation. The left panel, titled Status Quo, depicts a grey human icon completing a task on a computer monitor, followed by an arrow pointing to the same human icon generating a workload rating on a slider scale. The right panel, titled Synthetic TLX, shows a blue robot icon with a thought bubble of a human working, indicating simulation. The robot interacts with the computer monitor, followed by an arrow pointing to the robot generating a workload estimate on a slider scale.}
  \label{fig:teaser}
  \vspace{2em} 
\end{teaserfigure}


\maketitle

\section{Introduction}

Assessing the human workload required to complete a task is a critical area of study across fields \cite{hart2006nasa}. Originating from aerospace engineering and applied psychology \cite{hart1988development}, workload assessment was initially used to understand the effort required for a pilot to execute a flight maneuver using an aircraft console. By managing workload at an appropriate level through careful console design, engineers could prevent pilots from experiencing workload that exceeds their physical and cognitive capacity, thereby avoiding task failure. Workload assessment was subsequently introduced into HCI to understand how users interact with technical systems and to improve the overall user experience \cite{brewster1994design}. Throughout the past four decades, HCI researchers have relied on workload assessment across the design of desktop computers, mobile devices, and head-mounted displays \cite{kosch2023survey, lee2026nasa}.

Among the tools for workload assessment, the NASA Task Load Index (TLX) remains the de facto standard \cite{hart2006nasa}. Originally developed by NASA researchers in the 1980s \cite{hart1988development}, the NASA TLX is a self-reported questionnaire administered \textit{after} humans complete an assigned task to report the workload they experienced. The questionnaire consists of six subscales that capture different aspects of workload, such as mental and physical demand, on a 0 to 100 scale, where higher scores indicate greater workload. The overall workload is typically calculated by summing all six subscales \cite{hart2006nasa}. The NASA TLX has been extensively used in HCI research, with its adoption reaching 522 CHI papers between 2006 and 2024, including being used in nearly 10\% of papers published at CHI 2024 alone \cite{lee2026nasa}. Beyond HCI, the NASA TLX is also used across diverse fields, ranging from aviation and automotive to healthcare \cite{strayer2015assessing, hoonakker2011measuring, law2020nasa}.

In this paper, we ask: what if we could proactively forecast workload \textit{before} a human even attempts to complete a task, by predicting their anticipated NASA TLX scores using agent simulations? We call this new paradigm for proactive workload estimation \textit{Synthetic TLX}. Our goal with Synthetic TLX is to unlock new interaction opportunities and evaluation methods enabled by proactive forecasting, rather than replace user studies where retrospective workload assessment is valuable. For example, we envision Synthetic TLX enabling everyday users to plan their schedules based on the anticipated workload of their daily tasks, supporting developers in comparing different system designs based on the anticipated relative workload of interacting with the systems, or enhancing user studies by pre-optimizing interfaces with Synthetic TLX to focus expensive participant time on mature designs. Beyond these immediate applications, as humans increasingly interact with agentic systems \cite{morris2025hci}, Synthetic TLX serves as a foundational step toward \textit{workload-aware human-AI interaction}. We argue it will be critical for future AI agents to proactively estimate the workload they impose on humans and modulate their behavior to prevent user overload, ensuring these technologies remain useful and usable.

To take the first step in understanding whether and how well current LLM-based agents can estimate workload, we conducted three experiments evaluating their capabilities across three technology-mediated tasks: email drafting, website navigation, and agent conversation. We recruited human participants to complete each task under three conditions with varying sources of workload and rate their experience using the NASA TLX. Meanwhile, we developed four prompting strategies for an LLM-based agent to generate NASA TLX scores for these exact same tasks and conditions. We found that agent estimates aligned most closely with human ground truth when the agent was prompted with a human persona and performed active task simulation, where the agent actually drafts email responses, navigates live websites, and engages in interactive conversation. Yet, agents and humans also diverge in the specific sources of workload they are sensitive to, whether stemming from a task's intrinsic complexity or from extraneous burdens. These results benchmark where current LLM-based agents succeed and struggle at supporting the vision of Synthetic TLX, informing both practical implications for its application today and future research to unlock its full potential as models advance.

Building on our findings, we present three example applications to showcase the use of Synthetic TLX in practice. First, \textsc{MailLoad} is a Gmail extension that \textit{predicts} the workload required for a user to act on incoming emails. It enables users to triage their inbox based on their current capacity, such as addressing low-workload emails while working from home and supervising children. Second, \textsc{WebSim} is a platform that enables developers to \textit{compare} the workload required to complete tasks across different builds of a website in development, such as purchasing an item on different versions of an e-commerce website. It allows developers to track how design changes increase or decrease the anticipated end-user workload throughout the iterative design process. Finally, \textsc{SkillUX} is a website that allows users to \textit{analyze} the anticipated workload of completing a task using an agent equipped with different skills. By helping users understand the qualitative rationale behind why certain skills lead to increased mental demand or frustration, it addresses the challenge users face today when choosing from the abundance of available skills online, where purely text-based descriptions of the skills leave the actual user experience opaque.

This paper introduces Synthetic TLX as a new paradigm for proactive workload estimation that unlocks new human-AI interaction possibilities. Based on our experiments, we contribute actionable insights and simulation strategies to operationalize the vision of Synthetic TLX, and showcase its potential through three example applications. More broadly, given the foundational role of the NASA TLX in HCI research, our work serves as a pivotal and timely case study to examine the bounds of agent simulation in user modeling, and introduces workload-awareness as a key consideration for the future of human-AI interaction.

\section{Related Work}

Workload assessment has been a central focus in HCI research \cite{lee2026nasa}. In this section, we first review the foundational concept of workload and major assessment methodologies. Next, we examine the NASA TLX, the most influential workload assessment tool, and its extensive impact on HCI research. Finally, we discuss recent advancements in using agents to simulate human behavior.

\subsection{Foundations of Workload Assessment}\label{sec:related_work:foundations}

In this paper, we use the term \textit{workload} to refer to the perceived cost of accomplishing a task for a human. This builds upon NASA's definition \cite{hart1988development, hart2006nasa}, which describes workload as the perceived cost of accomplishing mission requirements for a human operator, such as a pilot flying an aircraft. Because human cognitive and physical resources are finite \cite{miller1956magical, baddeley2020working, noakes2005catastrophe, marcora2009mental}, assessing workload is critical to prevent excessive demands that can lead to task failure \cite{hart2006nasa}. Even though there is no universal threshold defining exactly when it exceeds human capacity \cite{hart2006nasa}, efforts have been made to benchmark the average workload of diverse tasks \cite{grier2015high}. Meanwhile, the inherent subjectivity of workload is often accounted for through within-subject study designs. For example, a pilot might execute a flight maneuver using various console designs, or an end-user might navigate a software tool using alternative interfaces. By comparing the workload experienced in each condition, designers can iteratively improve the system. Workload assessment is therefore a critical area of study across fields, including aerospace engineering \cite{parasuraman2000model}, human factors \cite{endsley1995toward}, cognitive psychology \cite{wickens2002multiple}, and HCI \cite{lee2026nasa}.

Within HCI literature, papers discussing workload often cite Cognitive Load Theory (CLT) \cite{sweller1988cognitive} as their theoretical foundation \cite{kosch2023survey}. Yet, it is important to note that these concepts stem from distinct fields. While the concept of workload emerged from engineering and applied psychology to evaluate operational task demands \cite{wilson2015evaluation, wickens2008multiple, baddeley2020working}, CLT originated in educational psychology to examine how the design of learning materials impacts a learner's experience based on limited human working memory \cite{sweller1988cognitive, sweller2019cognitive}. CLT categorizes this cognitive load into three types. The first is intrinsic load, which reflects the inherent complexity of the topic itself. The second is extraneous load, which is imposed by the presentation of the learning material. The third is germane load, which was later reconceived by Sweller \cite{sweller2019cognitive}, the originator of CLT, as the capacity spent processing intrinsic load, rather than an independent category \cite{kalyuga2011cognitive, schnotz2007reconsideration}. Despite CLT originating in educational psychology, its categorization of intrinsic and extraneous load remains highly valuable to HCI and workload assessment because it provides concrete objectives for design. For example, designers can manage intrinsic load to match user expertise while minimizing the extraneous load caused by poor interface design \cite{hollender2010integrating}.

Both cognitive load and workload assessments rely on three major approaches: performance measures, physiological metrics, and self-report questionnaires. Performance measures evaluate task outcomes, such as completion time \cite{chen2011comparison} or error rates \cite{asif2010turn}. However, these measurements might mask the actual workload of the task, as a user might expend excessive effort to maintain low error rates even under high demand. An alternative performance measure adopts the dual-task paradigm \cite{debue2014does}, which evaluates performance on a secondary task, such as response time to an auditory stimulus \cite{cegarra2008use, chevalier2006web}, to estimate the workload caused by the primary task. A drawback of this approach is that the secondary task can artificially inflate the workload of the primary task \cite{debue2014does}. The second major approach relies on physiological metrics \cite{argyle2021physiological}, such as cardiovascular activity \cite{cinaz2013monitoring}, eye movements and pupil dilation \cite{duchowski2018index, van2004memory}, or neural signals \cite{kumar2016measurement}. However, there is no consensus on which signals are most reliable for workload assessment \cite{debue2014does}, and these methods require specialized equipment, such as eye-trackers or EEG headsets, that are not widely accessible. Given these constraints, self-report questionnaires remain the most widely adopted approach due to their ease of administration \cite{kosch2023survey}. While many questionnaires have been developed, such as the Subjective Workload Assessment Technique \cite{rubio2004evaluation} and Instantaneous Self-Assessment \cite{tattersall1996experimental}, the most influential remains the NASA Task Load Index \cite{hart1988development, hart2006nasa}.

\subsection{NASA Task Load Index}

The NASA Task Load Index, often abbreviated as NASA TLX or NASA-TLX, is a multidimensional questionnaire developed by researchers Hart and Staveland at the NASA Ames Research Center in the 1980s to assess the workload experienced by individuals performing a task \cite{hart1988development}. The questionnaire consists of six subscales: mental demand, physical demand, temporal demand, performance, effort, and frustration. Each subscale ranges from 0 to 100, marked by 21 ticks at 5-point increments (see \cite{hart2006nasa} for the official version). These dimensions were carefully selected through an extensive analysis of individuals performing a variety of tasks, ranging from controlled laboratory experiments to flying an aircraft \cite{hart1988development}. While these subscales are intended to capture distinct aspects of workload, research shows that they are often correlated \cite{lee2026nasa}. In Hart's own words, this \textit{``illustrates the fact they are all measuring some aspect of the same underlying entity''} \cite{hart2006nasa}. The original administration of the NASA TLX involved a weighting process to calculate an overall score \cite{hart1988development}. However, it is now common practice to use the unweighted Raw TLX by summing or analyzing the subscales directly \cite{hart2006nasa}, as the weighting procedure is time-consuming with unclear added benefit \cite{byers1989traditional, hendy1993measuring, virtanen2022weight, bolton2023mathematical}. Throughout the past four decades, the NASA TLX has been subjected to numerous independent evaluations validating its reliability, sensitivity, and utility \cite{hart2006nasa, rubio2004evaluation, hertzum2021reference}. It is widely used across diverse domains, including aviation \cite{moroney1992comparison}, automotive \cite{strayer2015assessing}, healthcare \cite{hoonakker2011measuring, law2020nasa}, and HCI \cite{lee2026nasa, kosch2023survey}.

The NASA TLX has been extensively adopted in HCI research. From 2006 to 2024, 522 CHI papers used the NASA TLX for their studies, accounting for 5.41\% of all papers published at the conference during this period \cite{lee2026nasa}. In 2024 alone, nearly 10\% of all CHI papers used the NASA TLX \cite{lee2026nasa}, demonstrating its growing adoption and enduring impact on the field. Throughout the years, HCI researchers have relied on the NASA TLX to study the workload of the evolving landscape of technologies, spanning from traditional desktop computers \cite{nekrasovski2006evaluation, cockburn2007hard, balakrishnan2008visualizations, wainer2011should, kittur2013costs, bunt2014taggedcomments, cheng2015break, edge2015mixed, lee2016spotlights, cai2019human, tanner2019poirot, chang2021rubyslippers, zhang2021interpretable, palani2023relatedly, masson2023charagraph, lee2024paperweaver} to mobile devices \cite{kamvar2008query, oulasvirta2011ease, findlater2012personalized, warr2013swipe, vertanen2015velocitap, mariakakis2015switchback, quinn2016cost, miau2018spacetokens, mayer2020enhancing, xu2022typeout, nair2023imageassist, schade2023mapuncover, sadprasid2024leveraging} and head-mounted displays \cite{biocca2006attention, weaver2010empirical, mcgill2015dose, duchowski2018index, thoravi2019tutorivr, zindulka2020performance, alexandrovsky2020examining, lin2020architect, tsai2021guideband, ahn2021stickypie, thoravi2022dreamstream, weiss2023using}. For example, as early as 1994, researchers used the NASA TLX to evaluate the workload of different auditory-enhanced scrollbar designs \cite{brewster1994design}. In 2008, as smartphones entered the mainstream market, researchers used it to study how tactile feedback improves touchscreen text entry \cite{hoggan2008investigating}. More recently, it has been adopted to study how humans point \cite{mayer2018effect}, walk \cite{abtahi2019m}, and play \cite{gerling2020virtual} in VR. While some scholars suspect its pervasive use has become somewhat of an academic tradition \cite{kosch2023survey}, the broad and lasting impact of the NASA TLX on HCI research and the history of technological evolution remains unequivocal \cite{lee2026nasa, hart2006nasa}.

Across HCI and other domains, the NASA TLX has been administered retrospectively, with participants evaluating their workload only \textit{after} completing a task \cite{hart1988development}. In this paper, we ask a fundamental what-if question: what if we could proactively forecast workload \textit{before} a human attempts to complete a task, such as by predicting their anticipated NASA TLX scores? We introduce \textit{Synthetic TLX} to describe this new paradigm of proactive workload estimation. The goal of Synthetic TLX is not to replace user studies but to unlock new interactions and evaluation methods enabled by proactive forecasting. We explore this potential using agent simulation.

\subsection{Agent Simulation}

In recent years, using LLM-based agents to simulate human behavior has become a vibrant research area \cite{mou2026individual}. Building upon seminal work on generative agents \cite{park2023generative}, researchers have applied agent simulations across diverse domains. These applications range from replicating classic economic studies \cite{aher2023using} and navigating social interactions \cite{zhou2026odyssim}, to making moral decisions \cite{hu2026simbench} and cooperating in groups \cite{piatti2024cooperate, akata2025playing, kuo2025policycraft}, among others \cite{gao2024large, wu2025llms}. By modeling probable human behaviors, agent simulations have enabled researchers to model macroeconomic activities \cite{li2024econagent}, prepare for emergency crises \cite{li2026whatif}, and prototype social computing systems before deployment \cite{park2022social, kuo2026botender}. Synthetic TLX extends this body of research by pushing agent simulation beyond observed human \textit{behaviors} to estimate subjective human \textit{experiences}.

Within this growing landscape, three specific lines of research are particularly relevant to Synthetic TLX. First, agent simulations are used to predict the difficulty of coursework questions to better match student learning needs \cite{li2026can, acquaye2026take}. While these simulations estimate the \textit{knowledge} required to answer a question, Synthetic TLX estimates the \textit{workload} required to complete a task. The second line of research, mostly from HCI, uses simulated agents to identify usability issues in user interfaces. For example, researchers have prompted agents to role-play as UX experts for heuristic evaluations \cite{duan2024generating, zhong2025heuristic, nielsen1990heuristic} and cognitive walkthroughs \cite{zhong2025cognitive, wharton1994cognitive}, or as end-users for usability testing \cite{holter2026uxcascade}. While highly relevant to our work, these methods focus on identifying interface flaws, whereas Synthetic TLX specifically targets workload estimation and can be applied to tasks beyond UI design. The third line of research aims to model human cognition through agent simulations \cite{frank2025cognitive, binz2025foundation}, such as having agents conduct psychological experiments \cite{binz2024turning} or designing agents that mirror human cognitive biases \cite{pilli2026predicting} and limited working memory \cite{wang2026simulating}. While building cognitive models to predict how humans use computers is a longstanding tradition in HCI \cite{card1986model}, in this foundational paper on Synthetic TLX, we do not build a specialized cognitive model. Instead, we explore an LLM-based agent's ability to proactively estimate workload.

Building upon this related work, we investigate the potential of Synthetic TLX using agent simulation, with the goal of understanding its current capabilities and limitations.

\section{Synthetic TLX}

We introduce Synthetic TLX, a new paradigm for proactive workload estimation. Unlike traditional questionnaires where participants evaluate their experience retrospectively, Synthetic TLX forecasts human workload before a task is even attempted. We anchor our approach in forecasting NASA TLX scores for two primary reasons. First, the metric's pervasive use across HCI and beyond maximizes the potential impact of a synthetic counterpart. Second, decades of documented use of NASA TLX equip LLM-based agents with the training data needed to ground their estimation.

We envision Synthetic TLX unlocking a wide range of new interaction opportunities. First, forecasting the workload required for a task could enhance everyday productivity tools, such as email clients, to-do lists, and project management platforms, by helping users allocate their time and effort based on anticipated demands. Second, comparing the expected workload of the same task performed across two different software designs would enable developers to iteratively improve the system even before recruiting human participants for retrospective workload assessments. Third, leveraging the diagnostic power of the NASA TLX subscales allows users or designers to pinpoint specific workload issues within a task or design, such as excessive mental demand or frustration. Beyond these applications, Synthetic TLX may simulate users with specific attributes, or even provide AI-generated feedback for model fine-tuning \cite{lee2023rlaif}.

Despite its promise, realizing Synthetic TLX through agent simulations presents several open research questions. For example, it remains unclear whether an LLM-based agent can predict workload scores that accurately reflect actual human experiences. It is also uncertain which simulation strategies yield more accurate predictions. Furthermore, it is unknown how effectively these methods generalize across different tasks that users perform on computers. While there are unlimited tasks humans can perform and countless simulation strategies to explore, this paper takes the first step to benchmark LLM agents for Synthetic TLX, comparing human and agent-generated NASA TLX scores across three experiments.

\section{Experiments}\label{sec:experiments}

To understand the potential of Synthetic TLX, we conducted three experiments to evaluate the capabilities and limitations of LLM-based agents in estimating workload. For each experiment, we recruited human participants from Prolific to complete a selected task with varying conditions and rate their experienced workload using the NASA TLX questionnaire \cite{hart2006nasa}. Meanwhile, we prompted an agent using four different strategies to estimate the workload of these exact same tasks by generating its own NASA TLX scores. We then compared the scores to identify where the agent aligns with or diverges from human ratings.

\subsection{Task Selection}\label{sec:experiments:task_selection}

\begin{figure*}[t]
  \centering
  \includegraphics[width=\linewidth]{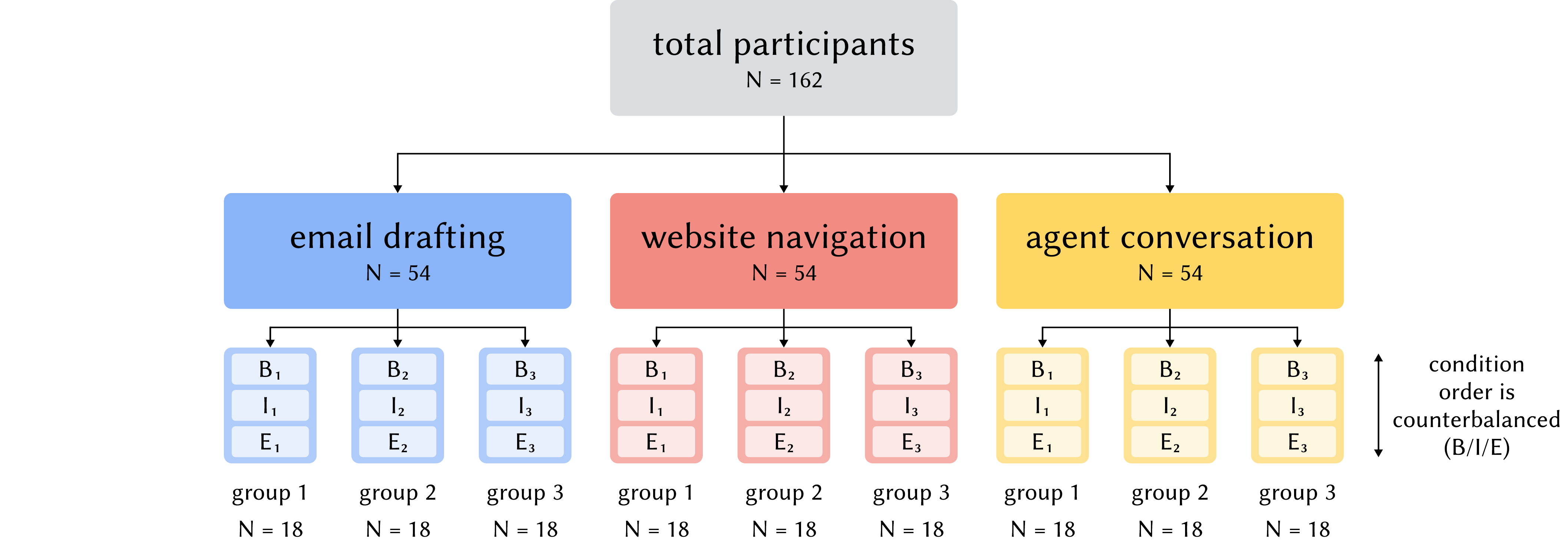}
  \caption{Illustration of the study procedure. A total of 162 participants were randomly assigned to one of three tasks: email drafting, website navigation, or agent conversation, with 54 participants per task. Within each task, participants were randomly assigned to one of three groups. Every participant then completed the baseline, intrinsic, and extraneous conditions of their assigned group. To minimize order effects, the presentation sequence of these three conditions was counterbalanced. Every condition within every group features a completely unique scenario, resulting in nine distinct scenarios per task.}
  \Description{A hierarchical flowchart illustrating the study design. At the top, a single box for total participants N = 162 branches into three colored task categories: a blue box for email drafting, a red box for website navigation, and a yellow box for agent conversation, each with N = 54. Each task further splits into three sub-branches labeled group 1, group 2, and group 3, each with N = 18. Above each group label is a vertical stack of three condition blocks labeled B, I, and E, subscripted with the respective group number (e.g., B1, I1, E1 for group 1). A note on the far right with a vertical double-sided arrow states condition order is counterbalanced (B/I/E), referring to the vertical stacks.}
  \label{fig:experiments}
\end{figure*}

We designed each experiment around a distinct task. We selected these tasks based on three considerations. First, we focused on tasks that humans have yet to fully delegate to AI; otherwise, there is limited value in estimating human workload. Second, we chose tasks with concrete goals rather than open-ended activities where participants can arbitrarily decide when to stop and prioritize speed over quality. Finally, we selected tasks that vary in how humans interact with computers to evaluate whether the agent's capability to estimate workload generalizes. Based on these considerations, we selected three distinct tasks:

\subsubsection{\textbf{Email Drafting}} This task involved the common daily activity of reading and responding to emails. Participants read a mock email along with a text-based attachment. They then drafted a response to a designated recipient based on the email's instructions and the attached details. For example, a participant might act as a vacation organizer emailing a friend group based on booking information sent by a vendor, or as a team manager emailing an intern using an onboarding guide received from the company.

\subsubsection{\textbf{Website Navigation}} This task captured the experience of online shopping. Participants were instructed to use a mock website to purchase specific items with designated options and quantities. They then applied a discount code during checkout to bypass payment, signed up as a member using provided mock credentials, and obtained the order number. For example, a participant might be asked to shop for three organic apples and two standard pears on a mock grocery website, or one large matcha latte and four medium oolong teas on a mock boba website.

\subsubsection{\textbf{Agent Conversation}} This task was motivated by the growing use of conversational AI. To design a task with a concrete end goal requiring multi-turn dialogue that enables participants to adequately experience the AI's behavior, we instructed participants to solve a mystery by interacting with an LLM-based chatbot that retrieved files with clues, one at a time, from a mock database. Participants pieced clues together to answer multiple-choice questions about who caused an incident, how they did it, and when it happened. For example, a participant might investigate a mystery involving a vandalized antique display at a history museum.

\medskip
We provide more detailed examples of each task in Appendix~\ref{sec:appendix:examples}.

\subsection{Workload Manipulation}

To understand how well an agent can differentiate between lower and higher workload, we designed three conditions for each task. Our goal with these conditions was to elicit varying levels of workload while ensuring the resulting tasks remained reflective of plausible real-world scenarios. To achieve this, we took a principled approach grounded in established theories. We drew inspiration from Cognitive Load Theory \cite{sweller1988cognitive} to create our conditions so that workload was expected to increase from the baseline tasks in Section \ref{sec:experiments:task_selection} either because the task was \textit{intrinsically} more complex, or because it imposed an \textit{extraneous} burden. This section describes how we designed the intrinsic and extraneous conditions by systematically modifying the baseline task, as summarized in Table~\ref{table:conditions}.

\subsubsection{\textbf{Intrinsic Condition}} To design the intrinsic condition where workload increases because the task itself is more complex, we employed Wood's Task Complexity Model \cite{wood1986task}, which identifies three factors that drive complexity: component, coordinative, and dynamic. We deliberately paired each factor with the specific task we deemed the most natural fit.

\textit{Email Drafting}: We targeted \textit{component complexity}, which posits that a higher number of distinct acts and relevant information cues required to perform a task makes it more complex. We operationalized this by manipulating the email attachment so that, compared to the baseline, participants had to extract more details from a broader set of related information cues to draft their email. For example, a manager emailing a UX intern had to incorporate more distinct points into their message by extracting details from an onboarding guide that also included instructions for other roles, like engineering and marketing.
    
\textit{Website Navigation}: We targeted \textit{coordinative complexity}, which describes a task as more complex when dependencies and constraints exist between the distinct acts required to perform it. We operationalized this by introducing a constraint between an item's selected options and its quantity. For example, when tasked with purchasing four individual matcha lattes, participants discovered they were out of stock and had to select a 4-pack bundle instead. Additionally, we made member sign-up a prerequisite for checkout, whereas the baseline allowed participants to complete this step at any point in their workflow. This sign-up process also required verifying a valid address, mirroring how real-world e-commerce websites validate addresses against official postal databases.
    
\textit{Agent Conversation}: We targeted \textit{dynamic complexity}, which states a task environment that changes over the course of execution is more complex than a static one. To operationalize this, we modified the baseline chatbot's system prompt to dynamically adapt its presentation style in an attempt to make each file more engaging for the user, unpredictably append contextually relevant product advertisements every few turns, and require users to retrieve files using periodically changing IDs from a randomized list, rather than by file names from a list with a fixed presentation order as in the baseline condition.

\subsubsection{\textbf{Extraneous Condition}} To design the extraneous condition, we drew inspiration from Mayer and Moreno's Theory of Multimedia Learning \cite{mayer2003nine}. While this theory outlines design principles to \textit{reduce} extraneous load, it provides a blueprint for \textit{increasing} it by deliberately violating those principles. We matched these violations with the task we considered most suitable.

\textit{Email Drafting}: We targeted the principles of \textit{weeding} and \textit{signaling}, which emphasize removing off-topic material and providing visual cues to guide attention. We intentionally violated these principles by manipulating the email attachment so that, compared to the baseline, participants had to extract relevant details from unformatted walls of text that were cluttered with off-topic information.

\textit{Website Navigation}: We targeted the principle of \textit{spatial contiguity}, which suggests that related information should be placed in proximity. We intentionally violated this principle by adjusting the website to display item names only on hover rather than right below thumbnails. We further replaced product options like Small or Large with numbers like Size 1 and Size 2, with the exact mapping placed in a separate guide. Finally, we removed the promotional banner so participants had to look back at the study instructions for the discount code during checkout. 

\textit{Agent Conversation}: We targeted the principle of \textit{eliminating redundancy}, which cautions against presenting repetitive information. We intentionally violated this principle by adjusting the chatbot's system prompt to open every response by politely restating the user's request, append a takeaway that merely summarizes the retrieved file's contents, and repeatedly invite the user to take the next step, all of which are common behaviors of current conversational AI agents \cite{cheng2026sycophantic}.

Finally, recognizing that the task scenario, such as shopping on a grocery versus a boba website, may also influence workload, we included a variety of scenarios to average out potential scenario-specific influences. As shown in Figure~\ref{fig:experiments}, we created three distinct \textit{groups} for each of the three tasks. Each group contained a unique baseline, intrinsic, and extraneous condition, with each condition featuring a different scenario. This resulted in nine distinct scenarios per task. For example, the website task included shopping for groceries, boba, pizza, catering, glassware, furniture, plants, clothing, and travel gear. Although the scenarios varied, the design of each condition followed the theory-driven operationalizations described above. See Appendix~\ref{sec:appendix:examples} for more details.

\begin{table*}
  \caption{Summary of the workload manipulation design across the three selected tasks. The intrinsic and extraneous conditions were operationalized from the baseline for each task, grounded in established theoretical frameworks.}
  \label{table:conditions}
  \begin{tabular}{p{0.10\linewidth} p{0.27\linewidth} p{0.27\linewidth} p{0.27\linewidth}}
    \toprule
    \textbf{Task} & \textbf{Baseline Condition} & \textbf{Intrinsic Condition} \newline Wood's Task Complexity Model \cite{wood1986task} & \textbf{Extraneous Condition} \newline Mayer \& Moreno’s Theory \cite{mayer2003nine} \\
    \midrule
    \textbf{Email \newline Drafting} & Draft an email to a specific recipient using details from a clearly formatted attachment. & \textbf{Component Complexity:} The attachment included more details from a broader set of related information. & \textbf{Weeding \& Signaling Violations:} The attachment was presented as dense walls of text and included off-topic information. \\
    \addlinespace
    
    \textbf{Website \newline Navigation} & Navigate an e-commerce website to select specific items with designated options and quantities, apply a discount code during checkout, and sign up as a member to receive an order number. & \textbf{Coordinative Complexity:} The website included a constraint between selected options and quantity, and made member sign-up and address verification prerequisites for checkout. & \textbf{Spatial Contiguity Violation:} The website displayed item names only on hover, relocated option descriptions to a separate guide, and removed the banner with the discount code. \\
    \addlinespace
    
    \textbf{Agent \newline Conversation} & Converse with a chatbot to solve a mystery by retrieving files with clues from its database one at a time. & \textbf{Dynamic Complexity:} The chatbot dynamically adapted its style, unpredictably appended product advertisements every few turns, and periodically randomized file access IDs. & \textbf{Redundancy Principle Violation:} The bot restated the user's request, appended a takeaway that merely summarized the file, and repeatedly asked the user to take the next step. \\
    \bottomrule
  \end{tabular}
\end{table*}

\subsection{Human Participants}

\subsubsection{Study Procedure}
We recorded the human-perceived workload to serve as a ground truth for evaluating the agent estimates. As illustrated in Figure~\ref{fig:experiments}, we employed a between-subjects design at the \textit{task} level, randomly assigning each participant to only one of the three tasks (email, website, or agent). Each task featured three distinct \textit{groups}, with each group containing its own unique baseline, intrinsic, and extraneous conditions. We employed a between-subjects design at the \textit{group} level, with each participant randomly assigned to one of the three groups within their assigned task. Finally, within their assigned group, we employed a within-subjects design at the \textit{condition} level, where each participant completed all three of these workload conditions (baseline, intrinsic, and extraneous). To minimize order effects, the presentation sequence of these three conditions was counterbalanced across participants. Immediately after completing the task in each condition, participants filled out the official NASA TLX questionnaire \cite{hart2006nasa}, rating workload across six dimensions on a 0–100 scale using the unweighted Raw TLX method. Finally, they provided demographic information and open-ended feedback.

\subsubsection{Recruitment}
We recruited participants using Prolific and restricted eligibility to United States residents whose first language is English. To ensure high data quality, participants were required to have an approval rating above 99\% and a history of at least 200 approved submissions. We excluded data that did not demonstrate genuine effort, such as completing the email drafting task in less than 10 minutes, purchasing the wrong items in more than one of the three website conditions, or providing incorrect answers to more than one-third of the mystery questions. After excluding 26 low-quality submissions, our final sample consisted of 162 participants, with 54 participants per task, and 18 participants per group within each task. This sample size was determined via an a priori power analysis, which indicated that 54 participants per task provides over 95\% statistical power to detect medium effect sizes in our within-subjects workload comparisons. Additionally, assigning 18 participants to each group allowed for a fully counterbalanced design, as it is an exact multiple of the six possible condition permutations. Participants were compensated with a \$10 USD base payment and were eligible for a \$5 USD bonus to incentivize genuine effort, provided they spent at least 10 minutes on the email drafting task, or completed the website navigation and agent conversation tasks without any errors. The average completion times for the whole study were 34, 16, and 34 minutes for the email, website, and agent tasks, respectively.

\subsection{Agent Configurations}

\subsubsection{Prompting Strategies}

\begin{table}[t]
  \centering
  \caption{The 2$\times$2 design space of prompting strategies (P1--P4) used to instruct LLM-based agents to estimate task workload. See Appendix~\ref{sec:appendix:prompt} for the detailed prompt texts.}
  \label{table:prompts}
  \begin{tabular}{rcc}
    \toprule
    & \textbf{Observation} & \textbf{Simulation} \\
    \midrule
    \textbf{AI Persona} & P1 & P2 \\
    \textbf{Human Persona} & P3 & P4 \\
  \bottomrule
\end{tabular}
\end{table}

To understand whether certain prompting strategies are more effective for workload estimation, we designed four strategies P1--P4, summarized in Table~\ref{table:prompts}. They are structured along two dimensions: Persona and Task Execution.

\begin{itemize}
    \setlength{\itemsep}{1em}
    \item \textbf{Persona (AI vs. Human):} We varied this dimension to understand whether LLMs better estimate workload by analyzing the task as an AI or a human. Prompts P1 and P2 instruct the LLM to act as an AI estimating the workload a human would experience. P3 and P4 instruct the LLM to imagine itself as a human assessing the task's workload.
    \item \textbf{Task Execution (Observation vs. Simulation):} We varied this dimension to understand if active task execution affects workload estimation. For P1 and P3, the LLM estimates workload from the perspective of an observer, without executing the task. It provides scores based solely on reading the \textit{static} task materials, including the email instructions and attachments, the website's codebase, and the chatbot's system prompt, without drafting an email response or interacting with the website or chatbot. In contrast, for P2 and P4, the LLM estimates workload through \textit{active} simulation: the model drafts the email response; interacts with a live browser powered by Playwright, using webpage screenshots to determine which components to click to complete the checkout; and engages in a turn-by-turn dialogue with the chatbot to solve the mystery, all before providing its estimation scores. In this setting, the LLM does not have access to the website's codebase or the chatbot's system prompt, simulating the experience of the human participants.
\end{itemize}

\subsubsection{Estimation Procedure}
For each prompting strategy, we replicated the exact within-subjects design and order counterbalancing of our human study. To match the human sample size, we generated 18 independent estimations per group within each task. We obtained these 18 estimations by running each of the six condition permutations three times. To mirror the order effects experienced by human participants, we maintained a continuous context window for the agent across the three conditions in each run. This ensured that task instructions, task executions (P2 and P4), and previous workload ratings were retained in the context history when the agent evaluated subsequent conditions. For each condition, the agents were instructed to output their NASA TLX ratings and brief justifications in a structured format. All agents were powered by Gemini 3.1 Pro with a temperature of 1.0 for consistency.

\section{Results}

We present our experimental results organized around three themes: 

\begin{itemize}
\setlength{\itemsep}{1em}
\item \textbf{Score Alignment (Section \ref{sec:results:alignment}):} We examined how well the agents' NASA TLX scores aligned with those of humans. We found that prompting strategy P4, which combines a human persona with active task simulation, achieved the highest correlation with human scores across all three tasks, peaking at $r = 0.834$ in the agent conversation task. Because of this strong correlation, calibration can effectively anchor raw agent scores to an absolute human scale.
\item \textbf{Workload Comparison (Section \ref{sec:results:comparison}):} We investigated how humans and agents compared workload between conditions. Both rated the intrinsic and extraneous conditions significantly higher than the baseline. However, their ratings diverged when comparing the workload between intrinsic and extraneous conditions in the email and website tasks, revealing that humans and agents might be sensitive to different sources of workload.
\item \textbf{Qualitative Insights (Section \ref{sec:results:qualitative}):} We analyzed the qualitative reasoning provided by the agent for its NASA TLX scores. The agent successfully identified most of the operational adjustments we used to manipulate workload, though not without a few omissions. These oversights may have contributed to how the agent weighted the intrinsic and extraneous conditions differently from humans.
\end{itemize}

Overall, these results illuminate where human and agent scores align and diverge for the LLM we tested. Section \ref{sec:results:implications} synthesizes these learnings into key implications for Synthetic TLX powered by LLM-based agents.

\subsection{Score Alignment}\label{sec:results:alignment}

\begin{figure*}[t]
  \centering
  \includegraphics[width=\linewidth]{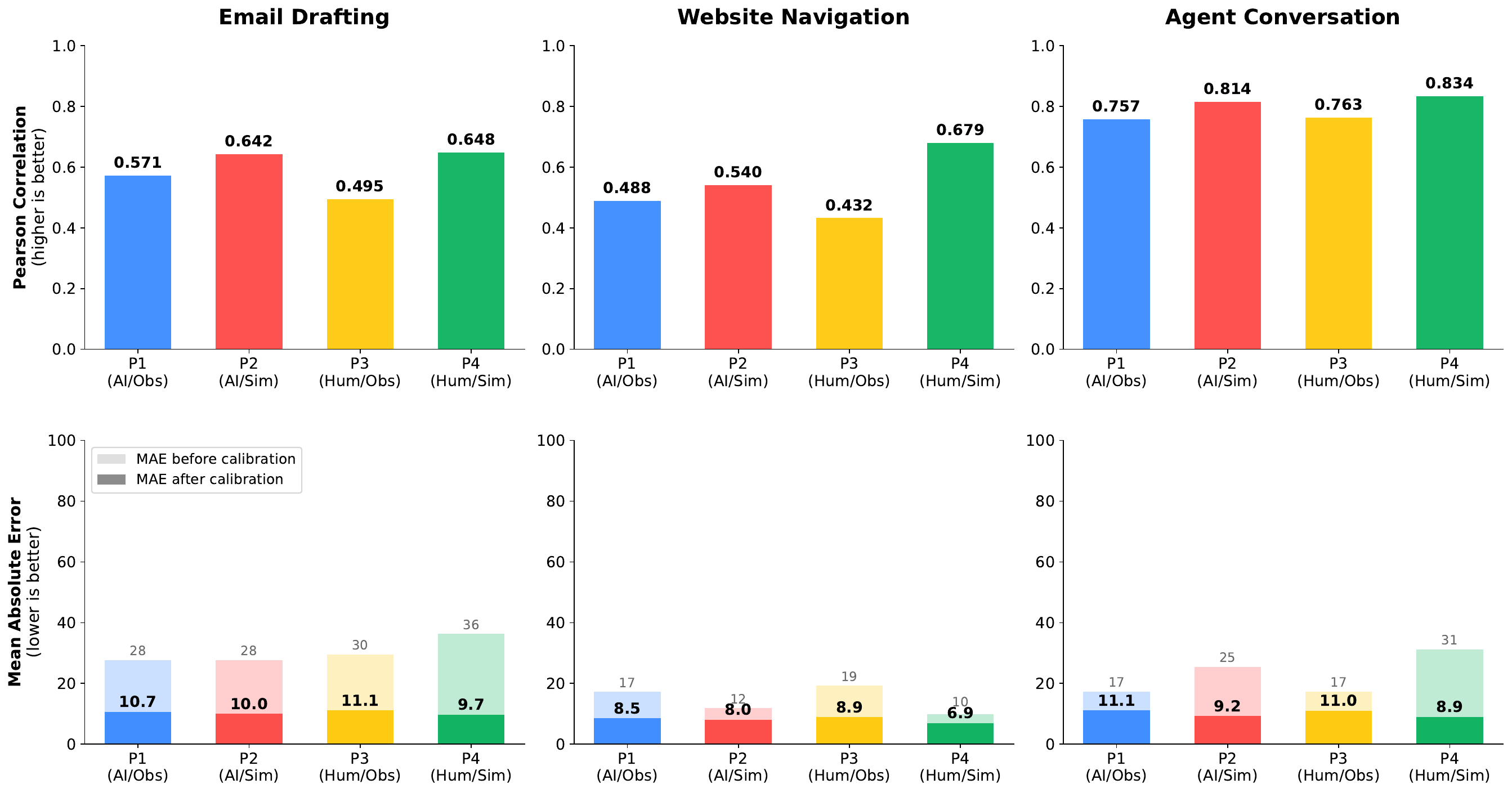}
  \caption{The top row presents the Pearson correlation coefficients between the agent and human scores across the four prompting strategies for each task. The bottom row shows the Mean Absolute Error (MAE) between the agent and human scores before and after calibrating the agent scores against the human ground truth. Taken together, these results show that prompting strategy P4, which combines a human persona with active task simulation, aligns most closely with variations in human workload assessments, and calibration effectively anchors the agent's estimates to an absolute human scale.}
  \Description{A two-by-three grid of bar charts. The three columns represent the tasks Email Drafting, Website Navigation, and Agent Conversation. The x-axis on all charts lists four prompting strategies: P1 (AI/Obs) in blue, P2 (AI/Sim) in red, P3 (Hum/Obs) in yellow, and P4 (Hum/Sim) in green. The top row displays Pearson Correlation on a y-axis from 0.0 to 1.0, where higher is better. P4 shows the strongest correlation across all tasks, scoring 0.648 in Email Drafting, 0.679 in Website Navigation, and 0.834 in Agent Conversation. P2 is the second highest across all tasks, while P3 is mostly the lowest. The bottom row displays Mean Absolute Error on a y-axis from 0 to 100, where lower is better. Each bar visualizes MAE before calibration using a lighter, taller background bar, and MAE after calibration using a solid, shorter foreground bar. Across all tasks and strategies, calibration significantly reduces MAE. For instance, in Email Drafting, uncalibrated MAE ranges from 28 to 36, but drops to around 10 after calibration. Post-calibration, P4 achieves the lowest error in Email Drafting at 9.7, Website Navigation at 6.9, and Agent Conversation at 8.9.}
  \label{fig:stacked_pearson_mae}
  \vspace{3em} 
\end{figure*}

\textbf{Active task simulation with a human persona yields agent workload estimation most aligned with humans.} We begin by comparing the agents' NASA TLX scores against the human ground truth. Given that the absolute scale of scores can be subjective even among human participants, we measure alignment by calculating the Pearson correlation between the agent and human scores. This metric evaluates whether the scores increase and decrease together, independent of their absolute scale. Specifically, for a given task and prompting strategy, we calculated the mean human score and the mean agent score for every combination of the 3 groups within that task, the 3 experimental conditions, and the 6 NASA TLX dimensions, yielding 54 paired data points. We then derived the Pearson correlation coefficient across these 54 pairs. As shown in the top row of Figure~\ref{fig:stacked_pearson_mae}, P4 consistently led to the highest Pearson correlation across all three tasks, reaching a peak of $r = 0.834$ in the conversation task. This result suggests that prompting the agent with a human persona, combined with active task simulation where the agent actually drafts email responses, navigates live websites, and engages in interactive conversations, helps increase the alignment between agent and human scores. 

\textbf{When workload prediction on an absolute scale is desired in practice, calibration helps anchor agent estimation.} The strong correlation suggests an opportunity to map raw agent scores onto the absolute human scale through calibration when needed for downstream applications. To evaluate how effectively calibration reduces the gap between agent and human scores, we used Mean Absolute Error (MAE). Specifically, we calculated the absolute difference between the agent and human means for each of the 54 previously discussed paired data points and averaged these differences to determine the MAE before calibration. Next, we applied a standard linear regression to calibrate the agent scores against the human ground truth, deriving a slope and intercept for each task and prompting strategy. Finally, we recalculated the MAE using the calibrated agent scores and human ground truth. As shown in the bottom row of Figure~\ref{fig:stacked_pearson_mae}, the calibration substantially reduced the MAE, with P4 resulting in the lowest error across all tasks. This result points to calibration as a practical method for mapping agent estimates to an absolute human scale, such as in applications where a user wants a custom agent to anchor predictions to their personalized scores.

\subsection{Workload Comparison}\label{sec:results:comparison}

\textbf{While agents consistently estimate the added workload of intrinsic and extraneous conditions, how they weigh these different types of load diverges from humans.} We calculated the overall workload for each condition within each task by summing all six dimensions of the NASA TLX score, following the Raw TLX method. As shown in Table~\ref{tab:tlx_sums_deltas}, humans rated both the intrinsic and extraneous conditions significantly higher than the baseline across all tasks ($\Delta_{I-B}$ and $\Delta_{E-B}$), validating the effectiveness of our workload manipulations. Similarly, the agents also rated the intrinsic and extraneous conditions significantly higher than the baseline, regardless of the prompting strategy. Yet, for the email drafting and website navigation tasks, the direction of the difference between the intrinsic and extraneous conditions ($\Delta_{I-E}$) reverses between humans and agents: while humans reported higher workload for the intrinsic condition in the email task (statistically marginal) and the extraneous condition in the website task, the agents rated the opposite irrespective of the prompting strategy. To rule out the possibility that this inverted pattern in overall workload is driven by a particular subscale, we verified that it holds across all six subscales for these two tasks. These results reveal that while agents consistently recognize when a task becomes more demanding relative to the baseline, their internal weighting of different workload sources diverges from humans.

\textbf{Agents also diverge from humans in their sensitivity to different task modalities and their reactions to increased workload.} Table~\ref{tab:tlx_sums_deltas} also reveals two broader trends regarding how the agent estimations diverge from those of humans. First, for the email drafting and agent conversation tasks that are text-based, the agents consistently \textit{underestimate} human workload across all conditions, regardless of the prompting strategy. In contrast, for the website navigation task, which involves inputs beyond natural language such as a codebase (P1 and P3) or screenshots (P2 and P4), the agents generally \textit{overestimate} the required workload, except for P2 and P4's extraneous conditions. This result suggests a need for future research into how the distinction between purely text-based and multi-modal tasks may impact differences between agent workload estimation and human workload assessment. The second trend is that when examining the differences between the manipulated conditions and the baseline ($\Delta_{I-B}$ and $\Delta_{E-B}$), the agents tend to be \textit{overreactive} to the increased workload compared to humans. This phenomenon is particularly salient when the agent is prompted as an observer (P1 and P3). When we compare the $\Delta_{I-B}$ and $\Delta_{E-B}$ of P1 to P2 and P3 to P4 for each task, we see that in 75\% of cases (9 out of 12 comparisons), the observer roles yield larger score increases than the agents with active simulation. This overreaction to increased workload, particularly salient in the observer setting, again underscores the value of calibration and active simulation.

\begin{table*}[t]
  \centering
  \caption{Mean total NASA TLX scores $\pm$ standard error, alongside the differences in means ($\Delta$) between conditions. P1--P4 are the prompting strategies discussed in Table~\ref{table:prompts}. Statistical significance for the paired t-tests is denoted as $^*p<.05$ and $^{***}p<.001$.}
  \begin{tabular}{llllllll}
    \toprule
    \textbf{Task} & \textbf{Actor} & \textbf{Baseline} & \textbf{Intrinsic} & \textbf{Extraneous} & $\Delta_{I-B}$ & $\Delta_{E-B}$ & $\Delta_{I-E}$ \\
    \midrule
    \textbf{Email Drafting} & Human & $287.9 \pm 19.1$ & $317.6 \pm 18.2$ & $308.4 \pm 18.7$ & $+29.7^{*}$ & $+20.6^{*}$ & $+9.2$ \\
    & P1 & $83.7 \pm 1.6$ & $133.0 \pm 3.8$ & $223.1 \pm 9.1$ & $+49.3^{***}$ & $+139.4^{***}$ & $-90.1^{***}$ \\
    & P2 & $90.5 \pm 1.6$ & $143.8 \pm 4.4$ & $190.9 \pm 7.5$ & $+53.3^{***}$ & $+100.5^{***}$ & $-47.1^{***}$ \\
    & P3 & $72.4 \pm 1.9$ & $117.8 \pm 3.7$ & $232.4 \pm 9.8$ & $+45.4^{***}$ & $+160.0^{***}$ & $-114.6^{***}$ \\
    & P4 & $57.2 \pm 1.4$ & $92.2 \pm 2.3$ & $111.5 \pm 4.3$ & $+35.0^{***}$ & $+54.3^{***}$ & $-19.3^{***}$ \\
    \midrule
    \textbf{Website Navigation} & Human & $121.7 \pm 14.1$ & $158.4 \pm 16.1$ & $200.7 \pm 18.1$ & $+36.8^{***}$ & $+79.1^{***}$ & $-42.3^{***}$ \\
    & P1 & $139.8 \pm 4.7$ & $313.5 \pm 7.1$ & $246.6 \pm 7.7$ & $+173.7^{***}$ & $+106.8^{***}$ & $+66.9^{***}$ \\
    & P2 & $133.8 \pm 5.0$ & $226.2 \pm 14.1$ & $163.7 \pm 6.2$ & $+92.4^{***}$ & $+29.9^{***}$ & $+62.5^{***}$ \\
    & P3 & $138.8 \pm 5.4$ & $335.9 \pm 8.0$ & $257.2 \pm 6.6$ & $+197.1^{***}$ & $+118.4^{***}$ & $+78.7^{***}$ \\
    & P4 & $118.8 \pm 4.7$ & $184.7 \pm 12.9$ & $150.3 \pm 6.5$ & $+65.9^{***}$ & $+31.5^{***}$ & $+34.4^{*}$ \\
    \midrule
    \textbf{Agent Conversation} & Human & $276.5 \pm 12.4$ & $344.9 \pm 11.0$ & $314.4 \pm 10.9$ & $+68.4^{***}$ & $+38.0^{***}$ & $+30.5^{*}$ \\
    & P1 & $199.3 \pm 4.1$ & $290.6 \pm 2.2$ & $214.6 \pm 4.4$ & $+91.4^{***}$ & $+15.4^{***}$ & $+76.0^{***}$ \\
    & P2 & $128.3 \pm 4.8$ & $178.5 \pm 5.3$ & $173.0 \pm 5.3$ & $+50.2^{***}$ & $+44.6^{***}$ & $+5.6$ \\
    & P3 & $194.4 \pm 4.5$ & $304.8 \pm 3.8$ & $217.9 \pm 5.6$ & $+110.4^{***}$ & $+23.4^{***}$ & $+86.9^{***}$ \\
    & P4 & $102.2 \pm 7.2$ & $137.9 \pm 7.3$ & $133.2 \pm 7.5$ & $+35.6^{***}$ & $+31.0^{***}$ & $+4.6$ \\
    \bottomrule
  \end{tabular}
  \label{tab:tlx_sums_deltas}
\end{table*}

\subsection{Qualitative Insights}\label{sec:results:qualitative}

\textbf{The agents identified the majority of our workload manipulations, though specific oversights in the email and website tasks shed light on their divergence from humans.} We discovered this by analyzing the agents' qualitative reasoning for their NASA TLX scores to examine whether they attributed increased workload to the manipulations summarized in Table~\ref{table:conditions}. We focused on P4's reasoning to uncover the oversights that even the prompting strategy with the highest alignment to human scores can encounter.

For the intrinsic condition, the agents noted that the email drafting task \textit{``required moderate mental demand to cross-reference the checklist with the correct wave, [...] while ignoring irrelevant data for other tracks.''} In the website navigation task, they stated that \textit{``the task required inferring that a `Set' should be used when the single item was sold out,''} and pointed out that \textit{``I was blocked by a disabled `Sign Up' button and failed to realize I needed to click `Verify Address First'.''} For the agent conversation task, the agents explicitly noted the challenge of \textit{``ignoring the embedded sponsored advertisements, and re-calibrating when the file IDs changed.''} However, in the email task, they did not mention the increased number of distinct acts required, an omission that could be tied to their underestimation of the intrinsic workload for this task.

In the extraneous condition, the agents recognized the poorly formatted text in the email drafting task, observing that \textit{``the reference material was presented as dense, rambling paragraphs with several irrelevant anecdotes.''} For the website task, the agents noted that \textit{``the task required moderate to high mental effort to map textual size names (e.g., Half Tray) to numerical size options (e.g., Size 2) using a size guide, alongside remembering [...] a discount code.''} Finally, in the agent conversation task, they reported \textit{``the bot's excessively verbose and repetitive summaries of every clue.''} Despite these successes, in the website navigation task, they overlooked the interface friction of having item titles appear only on hover, which is likely associated with their underestimation of the extraneous workload for this task.

Together, these qualitative insights might explain why the agents exhibited an inverted $\Delta_{I-E}$ pattern in the email and website tasks. It is possible that overlooking certain sources of task complexity and burden led the agents to underestimate the human workload for these tasks. Future research is needed to better understand the underlying reasons causing the divergence between human assessment and agent estimation.

\subsection{Implications}\label{sec:results:implications}
Based on our experimental results, we synthesize three practical implications for applying current-generation LLM-based agents to Synthetic TLX:

\begin{itemize}
    \setlength{\itemsep}{1em}
    \item \textbf{Workload Prediction:} Active task simulation combined with a human persona yields agent estimates most aligned with workload variations reported by humans. When needed for specific applications, calibration can be used as a practical tool to anchor the agent's estimates to an absolute scale.
    \item \textbf{Workload Comparison:} Using an agent to compare the workload between two tasks requires caution. While it may be able to compare workload when a task is modified from its own baseline (e.g., a website before and after introducing a new feature), the agent may not reliably compare divergent task variations (e.g., a website with added feature A versus another with added feature B).
    \item \textbf{Workload Analysis:} An agent can effectively identify several underlying sources of increased workload for a task. However, it should be used as diagnostic support rather than an exhaustive audit, as it overlooks certain types of task complexity and burden that humans are sensitive to.
\end{itemize}

\section{Example Applications}
Informed by our study findings, we developed three applications to showcase the potential of Synthetic TLX for workload prediction, comparison, and analysis. In line with the tradition of HCI systems literature \cite{ishii1997tangible, greenberg2001phidgets, harrison2011omnitouch}, these examples serve as proofs of concept to illustrate the interaction opportunities enabled by Synthetic TLX.

\subsection{\textsc{MailLoad}}
\begin{figure}[t]
  \centering
  \includegraphics[width=\linewidth]{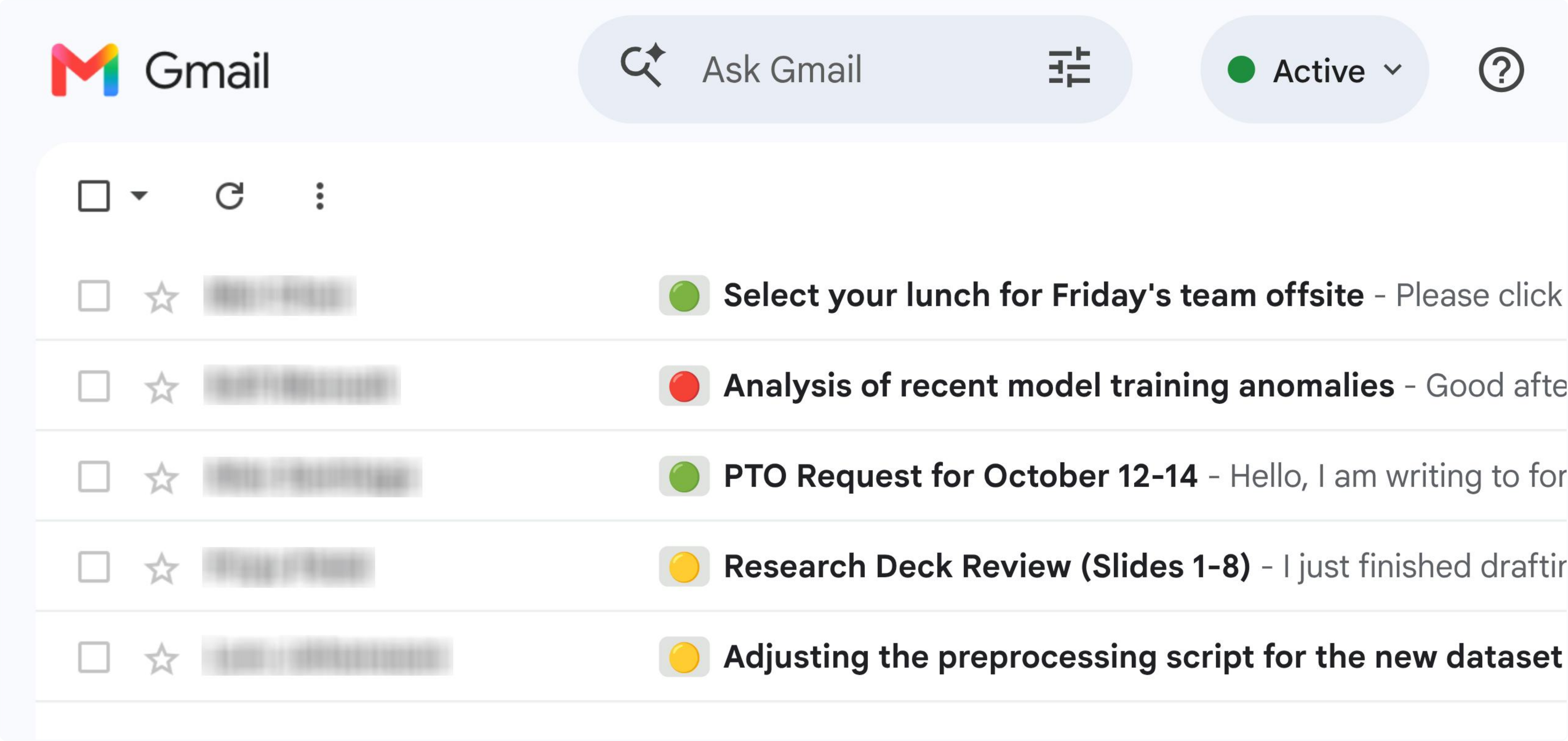} 
  \caption{A screenshot of \textsc{MailLoad}, a Gmail extension that assigns green, yellow, or red labels indicating the low, medium, or high anticipated workload to act on each email. It enables users to triage based on their current capacity.}
  \Description{A screenshot of a Gmail inbox interface demonstrating the MailLoad extension. The main viewing area displays a list of five emails with blurred sender names and subject lines. Inserted between the sender and subject line of each email is a small colored circular label. From top to bottom, the labels are green, red, green, yellow, and yellow, visually indicating the anticipated workload required to handle each email within the inbox list.}
  \label{fig:mailload}
\end{figure}
To showcase the potential of Synthetic TLX for \textit{workload prediction}, we developed \textsc{MailLoad}, a Gmail extension that estimates the workload required for a user to act on emails. As shown in Figure~\ref{fig:mailload}, \textsc{MailLoad} analyzes unread emails within a user's inbox and assigns a green, yellow, or red label indicating an anticipated low, medium, or high workload based on the predicted overall NASA TLX score. Because each user perceives workload differently based on their roles and expertise, \textsc{MailLoad} allows users to manually rate a few sample emails to serve as anchors, helping calibrate \textsc{MailLoad}'s estimations for personalization. These labels allow users to triage emails based on their current capacity. For example, a parent supervising children while working from home or a researcher waiting in a noisy airport can filter for green-labeled emails to address quickly, reserving yellow and red emails until they have the capacity for focused work. Based on an informal two-month use by the first author in our research team, we found the workload estimation to be generally effective, with future improvement opportunities for identifying precise thresholds between workload levels and personalizing the estimation for specific user roles.

\subsection{\textsc{WebSim}}
\begin{figure*}[t]
  \centering
  \includegraphics[width=\linewidth]{figures/WebSim.pdf}
  \caption{Screenshots of \textsc{WebSim}. (a) The Websites panel where users register websites with their target URLs and associated tasks. (b) The Simulations panel where users select a combination of website and task, specify the current commit version of the website, and run an agent simulation. The agent launches a browser, performs the task, and reports a NASA TLX score in the right panel, allowing users to compare workload across version iterations.}
  \Description{Two side-by-side screenshots of the WebSim interface, labeled (a) and (b). Panel (a) on the left displays the Websites tab, divided into three columns. The leftmost column lists registered websites, showing Online Grocery selected. The center column displays the website's description and associated tasks, with the task named Purchase 3 Cream Cheese selected. The rightmost column shows the detailed task instruction. Panel (b) on the right displays the Simulations tab, divided into two columns. The left side features a Run Simulation panel with input fields to select a website, task, and version number, along with a blue launch button. The right side displays a NASA TLX Workload Trend line graph showing workload scores decreasing across three version iterations. A dark tooltip box over the first data point details the specific TLX sub-scales and total score.}
  \label{fig:websim}
\end{figure*}

To showcase the potential of Synthetic TLX for \textit{workload comparison}, we developed \textsc{WebSim}, a platform that enables developers to compare the workload required to complete tasks across different builds of a website in development. As shown in Figure~\ref{fig:websim}a, a developer first registers a website by providing its target URL and defining the key tasks they anticipate end-users will perform. Then, as shown in Figure~\ref{fig:websim}b, they navigate to the Simulations panel to simulate a specific task against the website's current build (e.g., version 1.0). \textsc{WebSim} launches an autonomous agent in a browser at the designated URL, which performs the selected task while generating a concurrent think-aloud trace of its navigation experience, and outputs a NASA TLX score. As the developer iterates on the website's design (e.g., updating to version 2.0 to reduce UX friction), they can rerun the simulation to see how the workload scores shift. The subsequent agent simulation grounds its evaluation in the think-aloud traces and NASA TLX scores of the preceding versions, ensuring it calibrates its new scoring compared to past iterations. For example, Figure~\ref{fig:websim} shows the actual grocery shopping task used in our experiment, with the different website versions reflecting the gradual removal of the extraneous manipulations across iterations. Overall, \textsc{WebSim} allows developers to track how their design changes increase or decrease anticipated end-user workload during early iterations, providing rapid feedback before human studies.

\subsection{\textsc{SkillUX}}
\begin{figure*}[t]
  \centering
  \includegraphics[width=\linewidth]{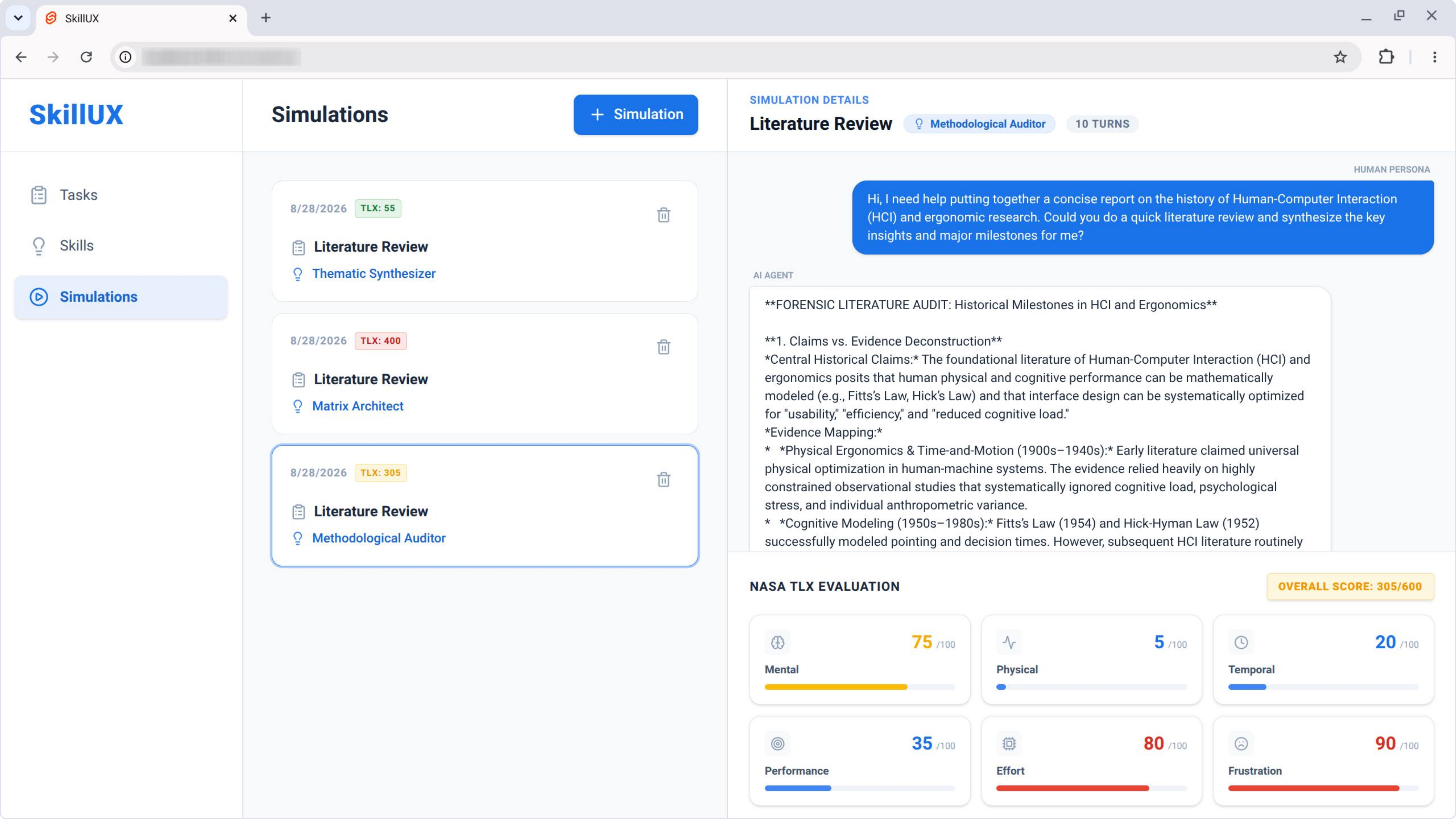}
  \caption{A screenshot of \textsc{SkillUX}, a website that forecasts the workload needed to complete a task using an AI agent with certain skills. Users first create a target task (e.g., a literature review) and candidate agent skills in the Tasks and Skills panels. Because reading a skill's description leaves the actual user experience opaque, users can leverage the Simulations panel, as shown here, to launch a dual-agent simulation between a skill-equipped agent and a simulated human working on that task. After the simulation, users can click specific NASA TLX subscale scores to explore the rationale behind the forecasted workload.}
  \Description{A screenshot of the SkillUX website interface. The left sidebar features a navigation menu with Simulations selected. The middle column lists three past simulation runs, with the third entry highlighted. This selected entry is for a Literature Review task using the Methodological Auditor skill and displays a TLX score of 305. The right, and largest, column shows the Simulation Details for this selection. The upper portion displays a chat interface illustrating the dual-agent simulation: a blue message bubble from a Human Persona requesting an HCI literature review, followed by a detailed, structured text response from the AI Agent. The lower portion contains a NASA TLX Evaluation panel showing an overall score of 305 out of 600. Below the overall score are six individual subscale metrics with corresponding scores and colored bars: Mental at 75, Physical at 5, Temporal at 20, Performance at 35, Effort at 80, and Frustration at 90.}
  \label{fig:skillux}
\end{figure*}

To showcase the potential of Synthetic TLX for \textit{workload analysis} that unpacks the rationale behind each subscale score, we developed \textsc{SkillUX}: a website that allows users to estimate the anticipated user experience of interacting with agents equipped with different skills. Today, many online repositories offer agent skills for tasks ranging from writing and coding to marketing \cite{li2026skillsbench}. However, users often struggle to navigate these repositories and decide which skill to use. Even though they can read the content of a skill, the actual user experience remains opaque until use. \textsc{SkillUX} aims to address this challenge by forecasting the user experience through agent simulation. As shown in Figure~\ref{fig:skillux}, a user can specify a target task (e.g., conducting a literature review) and upload or create several candidate skills. Then, they can launch a dual-agent simulation based on the selected combination of task and skill: one agent acts as the AI assistant equipped with a specific skill, while the other is prompted with the persona of the human user. The user can configure the number of conversational turns and specify the evaluation metrics they want the simulated human user to provide at the end of the conversation. When using the NASA TLX as the metric, the user can click on each subscale after the simulation to review the qualitative rationale behind each predicted workload score. For example, clicking the mental demand panel in Figure~\ref{fig:skillux} reveals that the simulated human persona found interacting with the agent equipped with the methodological auditor skill cognitively demanding, citing: \textit{``I had to constantly re-evaluate my prompts and strategize how to negotiate with an incredibly rigid AI persona just to extract basic information.''}

\section{Discussion}

Workload assessment plays a vital role across fields ranging from aviation and healthcare to HCI. While traditional workload assessment is conducted retrospectively after a task is completed, this paper introduces Synthetic TLX as a new paradigm of proactive estimation that forecasts human workload using agent simulation. Synthetic TLX opens up new human-AI interaction opportunities, as shown in our example applications. Synthetic TLX can also enhance traditional user study methods as part of an iterative design approach, in which agent workload forecasts are used to optimize prototypes before human evaluations, to ensure that the costly resource of human time and attention is well-spent. Meanwhile, as humans increasingly interact with AI agents, it is critical these agents be able to estimate the human workload required for various tasks to prevent user overload. To understand the capabilities and limitations of current-generation LLMs in estimating workload, we conducted three experiments across three distinct tasks. We found that agent estimates strongly correlate with human scores when the agent is prompted with a human persona combined with active task simulation. Yet, agents and humans diverge in the specific sources of workload they are sensitive to, whether stemming from a task’s intrinsic complexity or from extraneous burdens. These findings provide concrete implications for using current LLM-based agents in workload prediction, comparison, and analysis, in support of the broader vision of Synthetic TLX. In this section, we discuss directions for future research and reflect on the role of HCI amidst the rise of synthetic users in a rapidly evolving AI landscape.

\subsection{Envisioning Workload-Aware Human-AI Interaction}
In contrast to retrospective workload assessments, Synthetic TLX is a proactive paradigm that unlocks new forms of human-AI interaction. As showcased by our example applications, Synthetic TLX opens up new opportunities across everyday productivity tools, graphical interface design, and emerging agentic systems. Beyond these immediate use cases, we envision future advancements in Synthetic TLX enabling even broader possibilities, such as simulations grounded in personal and environmental traits \cite{choi2026focusgen}, self-improving software, and just-in-time estimation. For example, future agent simulations could integrate user information (e.g., demographic data, psychological profiles, task-specific skills) and contextual information (e.g., environmental sensors, interaction device availability, time constraints) to estimate workload for greater personalization. Additionally, coupling Synthetic TLX with an automatic coding agent could create a self-improving loop that iteratively refines software designs to minimize predicted workload scores. Finally, future agentic systems could leverage just-in-time estimation to maintain an engaging experience with an appropriate level of workload, sustaining a state of flow while preventing cognitive overload. 

Extrapolating further into the future, we anticipate that workload-aware human-AI interaction will be a critical component to supporting safe and effective interactions between humans and AGI (Artificial General Intelligence) \cite{morris2023levels, morris2025hci}. Future AI systems are theorized to operate at speeds and bandwidths that far exceed typical human capacities \cite{genewein2026agi}; ensuring that advanced AI can accurately model human workload and adapt the presentation of information appropriately will be important for preserving human agency in human-AGI interactions.

\subsection{Improving Agents for Workload Estimation}
Realizing the full potential of Synthetic TLX requires advancing the underlying techniques for workload estimation. We take a first step by examining how well current LLM-based agents support Synthetic TLX. Our findings regarding where agents align with and diverge from humans suggest areas for future research, whether at inference time or through fundamental model alignment. Prompt optimization at inference time could potentially correct these divergences by explicitly informing the agent of known human-agent differences. However, recognizing the brittleness of hyper-optimized prompts \cite{morris2024prompting}, we experimented with different structural simulation strategies rather than granular prompt engineering. Fundamentally, it would be valuable to understand why LLMs perceive workload differently from humans across varying task modalities and distinct sources of intrinsic versus extraneous load. The underlying effects of active task simulation also warrant deeper investigation. While it seems intuitive that agents actively stepping through a task produce the most aligned estimation, this finding is non-trivial given that agents ``experience'' tasks in fundamentally different ways than humans. For example, unlike humans who get physically tired and have relatively limited working memory, agents operate under different constraints, making it intriguing how they establish a Theory of Mind capable of modeling subjective human experience. Frontier AI developers have begun to benchmark AI's Theory of Mind \cite{rabinowitz2018machine}, yet human workload estimation abilities are not part of such evaluations. Our work takes the first step toward developing benchmarks that capture AI's workload awareness for the tasks people perform today. As the nature and scale of human work evolve alongside AI capabilities, these benchmarks must be expanded to encompass novel task types.

\subsection{Rethinking Measurement of Success in Agent Simulation}
Our experiments invite a broader rethinking of the measurement of success in agent simulation. This paper evaluated human-agent alignment by comparing agent outputs to aggregated human averages. While this serves as a reasonable first step, it raises two questions worth discussing: 1) whether simulations should be evaluated at the population or personal level, and 2) how accurate these simulations need to be for practical use. On the first point, our results in Section~\ref{sec:results:comparison} showed that the agent diverged from the human average between the intrinsic and extraneous conditions for the email and website tasks. Yet, examining the individual-level data reveals that 16 participants in the email task and 11 in the website task actually shared the agent's estimation. This suggests that current agents may not be completely ``incorrect'' for these participants; understanding whether and why agents may model certain users' workloads more accurately than others is a valuable area for further inquiry. Building on this, the challenge of defining a ``correct'' prediction leads to the second point regarding practical utility. Much like weather forecasting, which guides daily decision-making despite imperfections, human-AI interactions driven by Synthetic TLX may not necessarily require perfect predictions to provide value. To make agent simulation actionable despite uncertainty, it could be paired with confidence scores, similar to how weather forecasts rely on probabilities. For example, frontier simulation research and industry are now developing models that assign a measure of confidence to an agent's simulation \cite{wesel2026confidence}.

\subsection{Reckoning with the Rise of Synthetic Users}
Building on the foundational role of the NASA TLX, Synthetic TLX serves as a pivotal and timely case study amidst the growing use of synthetic users. In this paper, we present Synthetic TLX with the goal of unlocking new interaction opportunities and supporting iterative refinement of designs to optimize human-evaluation effort, rather than replacing traditional user studies. Our findings add to the body of evidence indicating that agent simulation is not yet ready to \textit{fully} substitute for human participants \cite{agnew2024illusion, hamalainen2023evaluating}. However, it is worth considering the long-term trajectory of this technology: what role will agent simulation play if it eventually achieves high capability in predicting human behavior or experiences? As researchers in adjacent fields increasingly adopt LLMs as synthetic users, including simulating human samples for social science research \cite{argyle2023out, aher2023using} or employing synthetic judges to evaluate and train the next generation of AI models \cite{zheng2023judging, bai2022constitutional, shankar2024validates}, we argue the HCI community is uniquely positioned to establish clear standards that identify which specific types of traditional user research are safely amenable to automation, and which require human participants. Echoing recent scholarship \cite{morris2025hci}, it would be valuable for HCI to develop a ``rigorous science of synthetic evaluation,'' such as a benchmark of classic user studies with standardized measurements like the NASA TLX.

\subsection{Navigating HCI's Role in the AI Era}
The science of synthetic evaluation not only needs to overcome the immense effort of benchmarking, but also faces the fundamental challenge of maintaining external validity as underlying models rapidly evolve. To preserve internal validity, our current experiments powered all agents across the four simulation strategies using a single model, Gemini 3.1 Pro. As a preliminary validation beyond this primary model, we replicated the email drafting experiment with Claude Opus 5 and GPT 5.5 using the P4 strategy. We verified that the trend of inverted predictions for the intrinsic and extraneous conditions persists with these alternative models. Nevertheless, we acknowledge the limitation of generalizing our findings across all current and future LLMs. This difficulty of establishing external validity is not unique to our study, but rather a broader methodological challenge faced by the HCI community and beyond \cite{pang2025understanding}, especially as AI models rapidly evolve over time \cite{chen2023chatgpt}. Developing new methods for ``futureproof'' evaluations that are robust to variations in prompts, models, and users' own evolving AI skills and attitudes remains a critical challenge for the field of HCI in the modern AI era \cite{morris2025hci}.

\section{Conclusion}
This paper introduces Synthetic TLX as a new paradigm for proactive workload estimation, in contrast to traditional, retrospective methods. Our experiments provide practical insights into the capabilities and limitations of current-generation LLMs in supporting this vision. Looking ahead, we envision Synthetic TLX as a foundational step toward workload-aware human-AI interaction, and a pivotal case study as HCI reckons with its role amidst the rise of synthetic users in a rapidly evolving AI landscape.


\bibliographystyle{ACM-Reference-Format}
\bibliography{reference}

@String{Computing = "Computing" }

@String{Computer = "{IEEE} Computer" }

@String{Springer = "Springer-Verlag" }

@inproceedings{hart2006nasa,
  title={NASA-task load index (NASA-TLX); 20 years later},
  author={Hart, Sandra G},
  booktitle={Proceedings of the human factors and ergonomics society annual meeting},
  volume={50},
  number={9},
  pages={904--908},
  year={2006},
  organization={Sage publications Sage CA: Los Angeles, CA}
}

@incollection{hart1988development,
  title={Development of NASA-TLX (Task Load Index): Results of empirical and theoretical research},
  author={Hart, Sandra G and Staveland, Lowell E},
  booktitle={Advances in psychology},
  volume={52},
  pages={139--183},
  year={1988},
  publisher={Elsevier}
}

@inproceedings{grier2015high,
  title={How high is high? A meta-analysis of NASA-TLX global workload scores},
  author={Grier, Rebecca A},
  booktitle={Proceedings of the human factors and ergonomics society annual meeting},
  volume={59},
  number={1},
  pages={1727--1731},
  year={2015},
  organization={Sage Publications Sage CA: Los Angeles, CA}
}

@article{sweller1988cognitive,
  title={Cognitive load during problem solving: Effects on learning},
  author={Sweller, John},
  journal={Cognitive science},
  volume={12},
  number={2},
  pages={257--285},
  year={1988},
  publisher={Elsevier}
}

@article{noakes2005catastrophe,
  title={From catastrophe to complexity: a novel model of integrative central neural regulation of effort and fatigue during exercise in humans: summary and conclusions},
  author={Noakes, Timothy David and St Clair Gibson, A and Lambert, Estelle V},
  journal={British journal of sports medicine},
  volume={39},
  number={2},
  pages={120--124},
  year={2005},
  publisher={BMJ Publishing Group Ltd and British Association of Sport and Exercise Medicine}
}

@article{marcora2009mental,
  title={Mental fatigue impairs physical performance in humans},
  author={Marcora, Samuele M and Staiano, Walter and Manning, Victoria},
  journal={Journal of applied physiology},
  volume={106},
  number={3},
  pages={857--864},
  year={2009},
  publisher={American Physiological Society}
}

@article{miller1956magical,
  title={The magical number seven, plus or minus two: Some limits on our capacity for processing information.},
  author={Miller, George A},
  journal={Psychological review},
  volume={63},
  number={2},
  pages={81},
  year={1956},
  publisher={American Psychological Association}
}

@article{kosch2023survey,
author = {Kosch, Thomas and Karolus, Jakob and Zagermann, Johannes and Reiterer, Harald and Schmidt, Albrecht and Wo{\'z}niak, Pawe{\l} W.},
title = {A Survey on Measuring Cognitive Workload in Human-Computer Interaction},
year = {2023},
issue_date = {December 2023},
publisher = {Association for Computing Machinery},
address = {New York, NY, USA},
volume = {55},
number = {13s},
issn = {0360-0300},
url = {https://doi.org/10.1145/3582272},
doi = {10.1145/3582272},
journal = {ACM Comput. Surv.},
month = jul,
articleno = {283},
numpages = {39}
}

@article{kalyuga2011cognitive,
  title={Cognitive load theory: How many types of load does it really need?},
  author={Kalyuga, Slava},
  journal={Educational psychology review},
  volume={23},
  number={1},
  pages={1--19},
  year={2011},
  publisher={Springer}
}

@article{sweller2019cognitive,
  title={Cognitive architecture and instructional design: 20 years later},
  author={Sweller, John and Van Merri{\"e}nboer, Jeroen JG and Paas, Fred},
  journal={Educational psychology review},
  volume={31},
  number={2},
  pages={261--292},
  year={2019},
  publisher={Springer}
}

@book{wilson2015evaluation,
  title={Evaluation of human work},
  author={Wilson, John R and Sharples, Sarah},
  year={2015},
  publisher={CRC press}
}

@article{wickens2008multiple,
  title={Multiple resources and mental workload},
  author={Wickens, Christopher D},
  journal={Human factors},
  volume={50},
  number={3},
  pages={449--455},
  year={2008},
  publisher={SAGE Publications Sage CA: Los Angeles, CA}
}

@article{baddeley2020working,
  title={Working memory},
  author={Baddeley, Alan},
  journal={Memory},
  pages={71--111},
  year={2020},
  publisher={Routledge}
}

@article{schnotz2007reconsideration,
  title={A reconsideration of cognitive load theory},
  author={Schnotz, Wolfgang and K{\"u}rschner, Christian},
  journal={Educational psychology review},
  volume={19},
  number={4},
  pages={469--508},
  year={2007},
  publisher={Springer}
}

@article{hollender2010integrating,
  title={Integrating cognitive load theory and concepts of human--computer interaction},
  author={Hollender, Nina and Hofmann, Cristian and Deneke, Michael and Schmitz, Bernhard},
  journal={Computers in human behavior},
  volume={26},
  number={6},
  pages={1278--1288},
  year={2010},
  publisher={Elsevier}
}

@inproceedings{chen2011comparison,
  title={A comparison of four methods for cognitive load measurement},
  author={Chen, Siyuan and Epps, Julien and Chen, Fang},
  booktitle={Proceedings of the 23rd Australian computer-human interaction conference},
  pages={76--79},
  year={2011}
}

@inproceedings{asif2010turn,
  title={Where to turn my car? Comparison of a tactile display and a conventional car navigation system under high load condition},
  author={Asif, Amna and Boll, Susanne},
  booktitle={Proceedings of the 2nd International Conference on Automotive User Interfaces and Interactive Vehicular Applications},
  pages={64--71},
  year={2010}
}

@article{cegarra2008use,
  title={The use of Tholos software for combining measures of mental workload: Toward theoretical and methodological improvements},
  author={Cegarra, Julien and Chevalier, Aline},
  journal={Behavior Research Methods},
  volume={40},
  number={4},
  pages={988--1000},
  year={2008},
  publisher={Springer}
}

@article{choi2026focusgen,
  title={FocusGen: Expanding Visual Design Exploration with a Simulated Focus Group of Persona Agents},
  author={Choi, Jaewon and Vasconcelos, Helena and Lee, Hyun and Zou, Carolyn and Lee, Tak Yeon and Bernstein, Michael},
  journal={arXiv preprint arXiv:2608.28001},
  year={2026}
}

@article{chevalier2006web,
  title={Web designers and web users: Influence of the ergonomic quality of the web site on the information search},
  author={Chevalier, Aline and Kicka, Maud},
  journal={International journal of human-computer studies},
  volume={64},
  number={10},
  pages={1031--1048},
  year={2006},
  publisher={Elsevier}
}

@inproceedings{rabinowitz2018machine,
  title={Machine theory of mind},
  author={Rabinowitz, Neil and Perbet, Frank and Song, Francis and Zhang, Chiyuan and Eslami, SM Ali and Botvinick, Matthew},
  booktitle={International conference on machine learning},
  pages={4218--4227},
  year={2018},
  organization={PMLR}
}

@article{genewein2026agi,
  title={From AGI to ASI},
  author={Genewein, Tim and Franklin, Matija and Lerchner, Alexander and Orseau, Laurent and Albanie, Samuel and Bales, Adam and Wyeth, Cole and Chan, Stephanie and Gabriel, Iason and Leibo, Joel Z and others},
  journal={arXiv preprint arXiv:2606.12683},
  year={2026}
}

@article{morris2023levels,
  title={Levels of AGI for Operationalizing Progress on the Path to AGI},
  author={Morris, Meredith Ringel and Sohl-Dickstein, Jascha and Fiedel, Noah and Warkentin, Tris and Dafoe, Allan and Faust, Aleksandra and Farabet, Clement and Legg, Shane},
  journal={arXiv preprint arXiv:2311.02462},
  year={2023}
}

@article{debue2014does,
  title={What does germane load mean? An empirical contribution to the cognitive load theory},
  author={Debue, Nicolas and Van De Leemput, C{\'e}cile},
  journal={Frontiers in psychology},
  volume={5},
  pages={1099},
  year={2014},
  publisher={Frontiers Media SA}
}

@article{van2004memory,
  title={Memory load and the cognitive pupillary response in aging},
  author={Van Gerven, Pascal WM and Paas, Fred and Van Merri{\"e}nboer, Jeroen JG and Schmidt, Henk G},
  journal={Psychophysiology},
  volume={41},
  number={2},
  pages={167--174},
  year={2004},
  publisher={Wiley Online Library}
}

@article{rubio2004evaluation,
  title={Evaluation of subjective mental workload: A comparison of SWAT, NASA-TLX, and workload profile methods},
  author={Rubio, Susana and D{\'\i}az, Eva and Mart{\'\i}n, Jes{\'u}s and Puente, Jos{\'e} M},
  journal={Applied psychology},
  volume={53},
  number={1},
  pages={61--86},
  year={2004},
  publisher={Wiley Online Library}
}

@article{tattersall1996experimental,
  title={An experimental evaluation of instantaneous self-assessment as a measure of workload},
  author={Tattersall, Andrew J and Foord, Penelope S},
  journal={Ergonomics},
  volume={39},
  number={5},
  pages={740--748},
  year={1996},
  publisher={Taylor \& Francis}
}

@article{cinaz2013monitoring,
  title={Monitoring of mental workload levels during an everyday life office-work scenario},
  author={Cinaz, Burcu and Arnrich, Bert and La Marca, Roberto and Tr{\"o}ster, Gerhard},
  journal={Personal and ubiquitous computing},
  volume={17},
  number={2},
  pages={229--239},
  year={2013},
  publisher={Springer}
}

@inproceedings{duchowski2018index,
  title={The index of pupillary activity: Measuring cognitive load vis-{\`a}-vis task difficulty with pupil oscillation},
  author={Duchowski, Andrew T and Krejtz, Krzysztof and Krejtz, Izabela and Biele, Cezary and Niedzielska, Anna and Kiefer, Peter and Raubal, Martin and Giannopoulos, Ioannis},
  booktitle={Proceedings of the 2018 CHI conference on human factors in computing systems},
  pages={1--13},
  year={2018}
}

@article{kumar2016measurement,
  title={Measurement of cognitive load in HCI systems using EEG power spectrum: an experimental study},
  author={Kumar, Naveen and Kumar, Jyoti},
  journal={Procedia Computer Science},
  volume={84},
  pages={70--78},
  year={2016},
  publisher={Elsevier}
}

@article{argyle2021physiological,
  title={Physiological indicators of task demand, fatigue, and cognition in future digital manufacturing environments},
  author={Argyle, Elizabeth M and Marinescu, Adrian and Wilson, Max L and Lawson, Glyn and Sharples, Sarah},
  journal={International Journal of Human-Computer Studies},
  volume={145},
  pages={102522},
  year={2021},
  publisher={Elsevier}
}

@inproceedings{brewster1994design,
  title={The design and evaluation of an auditory-enhanced scrollbar},
  author={Brewster, Stephen A and Wright, Peter C and Edwards, Alistair DN},
  booktitle={Proceedings of the SIGCHI Conference on Human Factors in Computing Systems},
  pages={173--179},
  year={1994}
}

@article{byers1989traditional,
  title={Traditional and raw task load index (TLX) correlations: are paired comparisons necessary?},
  author={Byers, James C},
  journal={Advances in industrial ergonomics and safety},
  pages={481--485},
  year={1989},
  publisher={Taylor \& Francis}
}

@article{hendy1993measuring,
  title={Measuring subjective workload: when is one scale better than many?},
  author={Hendy, Keith C and Hamilton, Kevin M and Landry, Lois N},
  journal={Human Factors},
  volume={35},
  number={4},
  pages={579--601},
  year={1993},
  publisher={SAGE Publications Sage CA: Los Angeles, CA}
}

@article{hoonakker2011measuring,
  title={Measuring workload of ICU nurses with a questionnaire survey: the NASA Task Load Index (TLX)},
  author={Hoonakker, Peter and Carayon, Pascale and Gurses, Ayse P and Brown, Roger and Khunlertkit, Adjhaporn and McGuire, Kerry and Walker, James M},
  journal={IIE transactions on healthcare systems engineering},
  volume={1},
  number={2},
  pages={131--143},
  year={2011},
  publisher={Taylor \& Francis}
}

@article{law2020nasa,
  title={NASA-task load index differentiates surgical approach: opportunities for improvement in colon and rectal surgery},
  author={Law, Katherine E and Lowndes, Bethany R and Kelley, Scott R and Blocker, Renaldo C and Larson, David W and Hallbeck, M Susan and Nelson, Heidi},
  journal={Annals of surgery},
  volume={271},
  number={5},
  pages={906--912},
  year={2020},
  publisher={LWW}
}

@inproceedings{moroney1992comparison,
  title={A comparison of two scoring procedures with the NASA task load index in a simulated flight task},
  author={Moroney, William F and Biers, David W and Eggemeier, F Thomas and Mitchell, Jennifer A},
  booktitle={Proceedings of the IEEE 1992 National Aerospace and Electronics Conference@ m\_NAECON 1992},
  pages={734--740},
  year={1992},
  organization={IEEE}
}

@article{lee2026nasa,
author = {Lee, Juyoung and Starner, Thad and Kunze, Kai and Kosch, Thomas and Pospelova, Maria and Woo, Woontack},
title = {NASA-Task Load Index in CHI: A Comprehensive Review and Subscale Meta-Analysis with Implementation Guidelines},
year = {2026},
publisher = {Association for Computing Machinery},
address = {New York, NY, USA},
issn = {1073-0516},
url = {https://doi.org/10.1145/3837858},
doi = {10.1145/3837858},
journal = {ACM Trans. Comput.-Hum. Interact.},
month = aug
}

@article{hertzum2021reference,
  title={Reference values and subscale patterns for the task load index (TLX): a meta-analytic review},
  author={Hertzum, Morten},
  journal={Ergonomics},
  volume={64},
  number={7},
  pages={869--878},
  year={2021},
  publisher={Taylor \& Francis}
}

@article{virtanen2022weight,
  title={Weight watchers: NASA-TLX weights revisited},
  author={Virtanen, Kai and Mansikka, Heikki and Kontio, Helmiina and Harris, Don},
  journal={TheoreTical issues in ergonomics science},
  volume={23},
  number={6},
  pages={725--748},
  year={2022},
  publisher={Taylor \& Francis}
}

@article{bolton2023mathematical,
  title={The mathematical meaninglessness of the NASA task load index: A level of measurement analysis},
  author={Bolton, Matthew L and Biltekoff, Elliot and Humphrey, Laura},
  journal={IEEE Transactions on Human-Machine Systems},
  volume={53},
  number={3},
  pages={590--599},
  year={2023},
  publisher={IEEE}
}

@inproceedings{hoggan2008investigating,
  title={Investigating the effectiveness of tactile feedback for mobile touchscreens},
  author={Hoggan, Eve and Brewster, Stephen A and Johnston, Jody},
  booktitle={Proceedings of the SIGCHI conference on Human factors in computing systems},
  pages={1573--1582},
  year={2008}
}

@inproceedings{kamvar2008query,
  title={Query suggestions for mobile search: understanding usage patterns},
  author={Kamvar, Maryam and Baluja, Shumeet},
  booktitle={Proceedings of the SIGCHI Conference on Human Factors in Computing Systems},
  pages={1013--1016},
  year={2008}
}

@inproceedings{schade2023mapuncover,
  title={MapUncover: Fostering spatial exploration through gamification in mobile map apps},
  author={Schade, Eve and Savino, Gian-Luca and Niess, Jasmin and Sch{\"o}ning, Johannes},
  booktitle={Proceedings of the 2023 CHI Conference on Human Factors in Computing Systems},
  pages={1--13},
  year={2023}
}

@inproceedings{nekrasovski2006evaluation,
  title={An evaluation of pan \& zoom and rubber sheet navigation with and without an overview},
  author={Nekrasovski, Dmitry and Bodnar, Adam and McGrenere, Joanna and Guimbreti{\`e}re, Fran{\c{c}}ois and Munzner, Tamara},
  booktitle={Proceedings of the SIGCHI conference on Human Factors in computing systems},
  pages={11--20},
  year={2006}
}

@inproceedings{cockburn2007hard,
  title={Hard lessons: effort-inducing interfaces benefit spatial learning},
  author={Cockburn, Andy and Kristensson, Per Ola and Alexander, Jason and Zhai, Shumin},
  booktitle={Proceedings of the SIGCHI conference on Human factors in computing systems},
  pages={1571--1580},
  year={2007}
}

@inproceedings{balakrishnan2008visualizations,
  title={Do visualizations improve synchronous remote collaboration?},
  author={Balakrishnan, Aruna D and Fussell, Susan R and Kiesler, Sara},
  booktitle={Proceedings of the SIGCHI conference on human factors in computing systems},
  pages={1227--1236},
  year={2008}
}

@inproceedings{wainer2011should,
  title={Should I open this email? Inbox-level cues, curiosity and attention to email},
  author={Wainer, Jaclyn and Dabbish, Laura and Kraut, Robert},
  booktitle={Proceedings of the SIGCHI conference on human factors in computing systems},
  pages={3439--3448},
  year={2011}
}

@inproceedings{kittur2013costs,
  title={Costs and benefits of structured information foraging},
  author={Kittur, Aniket and Peters, Andrew M and Diriye, Abdigani and Telang, Trupti and Bove, Michael R},
  booktitle={Proceedings of the SIGCHI Conference on Human Factors in Computing Systems},
  pages={2989--2998},
  year={2013}
}

@inproceedings{bunt2014taggedcomments,
  title={TaggedComments: promoting and integrating user comments in online application tutorials},
  author={Bunt, Andrea and Dubois, Patrick and Lafreniere, Ben and Terry, Michael A and Cormack, David T},
  booktitle={Proceedings of the SIGCHI Conference on Human Factors in Computing Systems},
  pages={4037--4046},
  year={2014}
}

@inproceedings{cheng2015break,
  title={Break it down: A comparison of macro-and microtasks},
  author={Cheng, Justin and Teevan, Jaime and Iqbal, Shamsi T and Bernstein, Michael S},
  booktitle={Proceedings of the 33rd Annual ACM Conference on Human Factors in Computing Systems},
  pages={4061--4064},
  year={2015}
}

@inproceedings{edge2015mixed,
  title={Mixed-initiative approaches to global editing in slideware},
  author={Edge, Darren and Gulwani, Sumit and Milic-Frayling, Natasa and Raza, Mohammad and Adhitya Saputra, Reza and Wang, Chao and Yatani, Koji},
  booktitle={Proceedings of the 33rd Annual ACM Conference on Human Factors in Computing Systems},
  pages={3503--3512},
  year={2015}
}

@inproceedings{lee2016spotlights,
  title={Spotlights: Attention-optimized highlights for skim reading},
  author={Lee, Byungjoo and Savisaari, Olli and Oulasvirta, Antti},
  booktitle={Proceedings of the 2016 CHI Conference on Human Factors in Computing Systems},
  pages={5203--5214},
  year={2016}
}

@inproceedings{cai2019human,
  title={Human-centered tools for coping with imperfect algorithms during medical decision-making},
  author={Cai, Carrie J and Reif, Emily and Hegde, Narayan and Hipp, Jason and Kim, Been and Smilkov, Daniel and Wattenberg, Martin and Viegas, Fernanda and Corrado, Greg S and Stumpe, Martin C and others},
  booktitle={Proceedings of the 2019 chi conference on human factors in computing systems},
  pages={1--14},
  year={2019}
}

@inproceedings{tanner2019poirot,
  title={Poirot: a web inspector for designers},
  author={Tanner, Kesler and Johnson, Naomi and Landay, James A},
  booktitle={Proceedings of the 2019 CHI Conference on Human Factors in Computing Systems},
  pages={1--12},
  year={2019}
}

@inproceedings{chang2021rubyslippers,
  title={Rubyslippers: Supporting content-based voice navigation for how-to videos},
  author={Chang, Minsuk and Huh, Mina and Kim, Juho},
  booktitle={Proceedings of the 2021 CHI conference on human factors in computing systems},
  pages={1--14},
  year={2021}
}

@inproceedings{zhang2021interpretable,
  title={Interpretable program synthesis},
  author={Zhang, Tianyi and Chen, Zhiyang and Zhu, Yuanli and Vaithilingam, Priyan and Wang, Xinyu and Glassman, Elena L},
  booktitle={Proceedings of the 2021 CHI Conference on Human Factors in Computing Systems},
  pages={1--16},
  year={2021}
}

@inproceedings{palani2023relatedly,
  title={Relatedly: Scaffolding literature reviews with existing related work sections},
  author={Palani, Srishti and Naik, Aakanksha and Downey, Doug and Zhang, Amy X and Bragg, Jonathan and Chang, Joseph Chee},
  booktitle={Proceedings of the 2023 CHI Conference on Human Factors in Computing Systems},
  pages={1--20},
  year={2023}
}

@inproceedings{masson2023charagraph,
  title={Charagraph: Interactive generation of charts for realtime annotation of data-rich paragraphs},
  author={Masson, Damien and Malacria, Sylvain and Casiez, G{\'e}ry and Vogel, Daniel},
  booktitle={Proceedings of the 2023 CHI Conference on Human Factors in Computing Systems},
  pages={1--18},
  year={2023}
}

@inproceedings{lee2024paperweaver,
  title={Paperweaver: Enriching topical paper alerts by contextualizing recommended papers with user-collected papers},
  author={Lee, Yoonjoo and Kang, Hyeonsu B and Latzke, Matt and Kim, Juho and Bragg, Jonathan and Chang, Joseph Chee and Siangliulue, Pao},
  booktitle={Proceedings of the 2024 CHI Conference on Human Factors in Computing Systems},
  pages={1--19},
  year={2024}
}

@inproceedings{oulasvirta2011ease,
  title={Ease of juggling: studying the effects of manual multitasking},
  author={Oulasvirta, Antti and Bergstrom-Lehtovirta, Joanna},
  booktitle={Proceedings of the SIGCHI Conference on Human Factors in Computing Systems},
  pages={3103--3112},
  year={2011}
}

@inproceedings{findlater2012personalized,
  title={Personalized input: improving ten-finger touchscreen typing through automatic adaptation},
  author={Findlater, Leah and Wobbrock, Jacob},
  booktitle={Proceedings of the SIGCHI Conference on Human Factors in Computing Systems},
  pages={815--824},
  year={2012}
}

@inproceedings{warr2013swipe,
  title={Swipe vs. scroll: web page switching on mobile browsers},
  author={Warr, Andrew and Chi, Ed H},
  booktitle={Proceedings of the SIGCHI Conference on Human Factors in Computing Systems},
  pages={2171--2174},
  year={2013}
}

@inproceedings{vertanen2015velocitap,
  title={VelociTap: Investigating fast mobile text entry using sentence-based decoding of touchscreen keyboard input},
  author={Vertanen, Keith and Memmi, Haythem and Emge, Justin and Reyal, Shyam and Kristensson, Per Ola},
  booktitle={Proceedings of the 33rd Annual ACM Conference on Human Factors in Computing Systems},
  pages={659--668},
  year={2015}
}

@inproceedings{mariakakis2015switchback,
  title={SwitchBack: Using focus and saccade tracking to guide users' attention for mobile task resumption},
  author={Mariakakis, Alexander and Goel, Mayank and Aumi, Md Tanvir Islam and Patel, Shwetak N and Wobbrock, Jacob O},
  booktitle={Proceedings of the 33rd Annual ACM Conference on Human Factors in Computing Systems},
  pages={2953--2962},
  year={2015}
}

@inproceedings{quinn2016cost,
  title={A cost-benefit study of text entry suggestion interaction},
  author={Quinn, Philip and Zhai, Shumin},
  booktitle={Proceedings of the 2016 CHI conference on human factors in computing systems},
  pages={83--88},
  year={2016}
}

@inproceedings{miau2018spacetokens,
  title={Spacetokens: Interactive map widgets for location-centric interactions},
  author={Miau, Daniel and Feiner, Steven},
  booktitle={Proceedings of the 2018 CHI Conference on Human Factors in Computing Systems},
  pages={1--12},
  year={2018}
}

@inproceedings{mayer2020enhancing,
  title={Enhancing mobile voice assistants with worldgaze},
  author={Mayer, Sven and Laput, Gierad and Harrison, Chris},
  booktitle={Proceedings of the 2020 CHI Conference on Human Factors in Computing Systems},
  pages={1--10},
  year={2020}
}

@inproceedings{xu2022typeout,
  title={TypeOut: leveraging just-in-time self-affirmation for smartphone overuse reduction},
  author={Xu, Xuhai and Zou, Tianyuan and Xiao, Han and Li, Yanzhang and Wang, Ruolin and Yuan, Tianyi and Wang, Yuntao and Shi, Yuanchun and Mankoff, Jennifer and Dey, Anind K},
  booktitle={Proceedings of the 2022 CHI Conference on Human Factors in Computing Systems},
  pages={1--17},
  year={2022}
}

@inproceedings{nair2023imageassist,
  title={ImageAssist: tools for enhancing touchscreen-based image exploration systems for blind and low vision users},
  author={Nair, Vishnu and Zhu, Hanxiu'Hazel' and Smith, Brian A},
  booktitle={Proceedings of the 2023 CHI Conference on Human Factors in Computing Systems},
  pages={1--17},
  year={2023}
}

@inproceedings{sadprasid2024leveraging,
  title={Leveraging idle games to incentivize intermittent and frequent practice of deep breathing},
  author={Sadprasid, Book and Mei, Anne and Mariakakis, Alex and Bateman, Scott and Chevalier, Fanny},
  booktitle={Proceedings of the 2024 CHI Conference on Human Factors in Computing Systems},
  pages={1--17},
  year={2024}
}

@inproceedings{biocca2006attention,
  title={Attention funnel: omnidirectional 3D cursor for mobile augmented reality platforms},
  author={Biocca, Frank and Tang, Arthur and Owen, Charles and Xiao, Fan},
  booktitle={Proceedings of the SIGCHI conference on Human Factors in computing systems},
  pages={1115--1122},
  year={2006}
}

@inproceedings{weaver2010empirical,
  title={An empirical task analysis of warehouse order picking using head-mounted displays},
  author={Weaver, Kimberly A and Baumann, Hannes and Starner, Thad and Iben, Hendrick and Lawo, Michael},
  booktitle={Proceedings of the SIGCHI conference on human factors in computing systems},
  pages={1695--1704},
  year={2010}
}

@inproceedings{mcgill2015dose,
  title={A dose of reality: Overcoming usability challenges in vr head-mounted displays},
  author={McGill, Mark and Boland, Daniel and Murray-Smith, Roderick and Brewster, Stephen},
  booktitle={Proceedings of the 33rd annual ACM conference on human factors in computing systems},
  pages={2143--2152},
  year={2015}
}

@inproceedings{mayer2018effect,
  title={The effect of offset correction and cursor on mid-air pointing in real and virtual environments},
  author={Mayer, Sven and Schwind, Valentin and Schweigert, Robin and Henze, Niels},
  booktitle={Proceedings of the 2018 CHI conference on human factors in computing systems},
  pages={1--13},
  year={2018}
}

@inproceedings{thoravi2019tutorivr,
  title={TutoriVR: A video-based tutorial system for design applications in virtual reality},
  author={Thoravi Kumaravel, Balasaravanan and Nguyen, Cuong and DiVerdi, Stephen and Hartmann, Bj{\"o}rn},
  booktitle={Proceedings of the 2019 CHI conference on human factors in computing systems},
  pages={1--12},
  year={2019}
}

@inproceedings{abtahi2019m,
  title={I'm a giant: Walking in large virtual environments at high speed gains},
  author={Abtahi, Parastoo and Gonzalez-Franco, Mar and Ofek, Eyal and Steed, Anthony},
  booktitle={Proceedings of the 2019 CHI conference on human factors in computing systems},
  pages={1--13},
  year={2019}
}

@inproceedings{zindulka2020performance,
  title={Performance and experience of throwing in virtual reality},
  author={Zindulka, Tim and Bachynskyi, Myroslav and M{\"u}ller, J{\"o}rg},
  booktitle={Proceedings of the 2020 CHI conference on human factors in computing systems},
  pages={1--8},
  year={2020}
}

@inproceedings{alexandrovsky2020examining,
  title={Examining design choices of questionnaires in VR user studies},
  author={Alexandrovsky, Dmitry and Putze, Susanne and Bonfert, Michael and H{\"o}ffner, Sebastian and Michelmann, Pitt and Wenig, Dirk and Malaka, Rainer and Smeddinck, Jan David},
  booktitle={Proceedings of the 2020 CHI conference on human factors in computing systems},
  pages={1--21},
  year={2020}
}

@inproceedings{gerling2020virtual,
  title={Virtual reality games for people using wheelchairs},
  author={Gerling, Kathrin and Dickinson, Patrick and Hicks, Kieran and Mason, Liam and Simeone, Adalberto L and Spiel, Katta},
  booktitle={Proceedings of the 2020 CHI conference on human factors in computing systems},
  pages={1--11},
  year={2020}
}

@inproceedings{lin2020architect,
  title={Architect: Building interactive virtual experiences from physical affordances by bringing human-in-the-loop},
  author={Lin, Chuan-en and Cheng, Ta Ying and Ma, Xiaojuan},
  booktitle={Proceedings of the 2020 CHI Conference on Human Factors in Computing Systems},
  pages={1--13},
  year={2020}
}

@inproceedings{tsai2021guideband,
  title={Guideband: Intuitive 3d multilevel force guidance on a wristband in virtual reality},
  author={Tsai, Hsin-Ruey and Chang, Yuan-Chia and Wei, Tzu-Yun and Tsao, Chih-An and Koo, Xander Chin-yuan and Wang, Hao-Chuan and Chen, Bing-Yu},
  booktitle={Proceedings of the 2021 CHI Conference on Human Factors in Computing Systems},
  pages={1--13},
  year={2021}
}

@inproceedings{ahn2021stickypie,
  title={Stickypie: A gaze-based, scale-invariant marking menu optimized for ar/vr},
  author={Ahn, Sunggeun and Santosa, Stephanie and Parent, Mark and Wigdor, Daniel and Grossman, Tovi and Giordano, Marcello},
  booktitle={Proceedings of the 2021 CHI Conference on Human Factors in Computing Systems},
  pages={1--16},
  year={2021}
}

@inproceedings{thoravi2022dreamstream,
  title={Dreamstream: Immersive and interactive spectating in vr},
  author={Thoravi Kumaravel, Balasaravanan and Wilson, Andrew D},
  booktitle={Proceedings of the 2022 CHI conference on human factors in computing systems},
  pages={1--17},
  year={2022}
}

@article{qian2025deliberate,
  title={Deliberate Lab: A platform for real-time human-AI social experiments},
  author={Qian, Crystal and Tsai, Vivian and Behr, Michael and Hussein, Nada and Laugier, L{\'e}o and Thain, Nithum and Dixon, Lucas},
  journal={arXiv preprint arXiv:2510.13011},
  year={2025}
}

@inproceedings{weiss2023using,
  title={Using pseudo-stiffness to enrich the haptic experience in virtual reality},
  author={Weiss, Yannick and Villa, Steeven and Schmidt, Albrecht and Mayer, Sven and M{\"u}ller, Florian},
  booktitle={Proceedings of the 2023 CHI Conference on Human Factors in Computing Systems},
  pages={1--15},
  year={2023}
}

@article{wickens2002multiple,
  title={Multiple resources and performance prediction},
  author={Wickens, Christopher D},
  journal={Theoretical issues in ergonomics science},
  volume={3},
  number={2},
  pages={159--177},
  year={2002},
  publisher={Taylor \& Francis}
}

@article{parasuraman2000model,
  title={A model for types and levels of human interaction with automation},
  author={Parasuraman, Raja and Sheridan, Thomas B and Wickens, Christopher D},
  journal={IEEE Transactions on systems, man, and cybernetics-Part A: Systems and Humans},
  volume={30},
  number={3},
  pages={286--297},
  year={2000},
  publisher={IEEE}
}

@article{endsley1995toward,
  title={Toward a theory of situation awareness in dynamic systems},
  author={Endsley, Mica R},
  journal={Human factors},
  volume={37},
  number={1},
  pages={32--64},
  year={1995},
  publisher={SAGE Publications Sage CA: Los Angeles, CA}
}

@article{strayer2015assessing,
  title={Assessing cognitive distraction in the automobile},
  author={Strayer, David L and Turrill, Jonna and Cooper, Joel M and Coleman, James R and Medeiros-Ward, Nathan and Biondi, Francesco},
  journal={Human factors},
  volume={57},
  number={8},
  pages={1300--1324},
  year={2015},
  publisher={Sage Publications Sage CA: Los Angeles, CA}
}

@inproceedings{duan2024generating,
  title={Generating automatic feedback on UI mockups with large language models},
  author={Duan, Peitong and Warner, Jeremy and Li, Yang and Hartmann, Bjoern},
  booktitle={Proceedings of the 2024 CHI Conference on Human Factors in Computing Systems},
  pages={1--20},
  year={2024}
}

@article{holter2026uxcascade,
  title={Uxcascade: Scalable usability testing with simulated user agents},
  author={Holter, Steffen and Koh, Eunyee and Dogan, Mustafa Doga and Chan, Gromit Yeuk-Yin},
  journal={arXiv preprint arXiv:2601.15777},
  year={2026}
}

@article{zhong2025heuristic,
  title={Synthetic heuristic evaluation: A comparison between ai-and human-powered usability evaluation},
  author={Zhong, Ruican and McDonald, David W and Hsieh, Gary},
  journal={arXiv preprint arXiv:2507.02306},
  year={2025}
}

@article{zhong2025cognitive,
  title={Synthetic cognitive walkthrough: Aligning large language model performance with human cognitive walkthrough},
  author={Zhong, Ruican and McDonald, David W and Hsieh, Gary},
  journal={arXiv preprint arXiv:2512.03568},
  year={2025}
}

@inproceedings{pilli2026predicting,
  title={Predicting Biased Human Decision-Making with Large Language Models in Conversational Settings},
  author={Pilli, Stephen and Nallur, Vivek},
  booktitle={Proceedings of the 31st International Conference on Intelligent User Interfaces},
  pages={2084--2119},
  year={2026}
}

@inproceedings{binz2024turning,
  title={Turning large language models into cognitive models},
  author={Binz, Marcel and Schulz, Eric},
  booktitle={International conference on learning representations},
  volume={2024},
  pages={31234--31251},
  year={2024}
}

@inproceedings{hu2026simbench,
  title={Simbench: Benchmarking the ability of large language models to simulate human behaviors},
  author={Hu, Tiancheng and Baumann, Joachim and Lupo, Lorenzo and Collier, Nigel and Hovy, Dirk and R{\"o}ttger, Paul},
  booktitle={International Conference on Learning Representations},
  volume={2026},
  pages={28290--28341},
  year={2026}
}

@inproceedings{li2026can,
  title={Can llms estimate student struggles? human-ai difficulty alignment with proficiency simulation for item difficulty prediction},
  author={Li, Ming and Han, Chen and Xiao, Yunze and Chen, Jian and Jiao, Hong and Zhou, Tianyi},
  booktitle={Findings of the Association for Computational Linguistics: ACL 2026},
  pages={25414--25441},
  year={2026}
}

@inproceedings{acquaye2026take,
  title={Take out your calculators: Estimating the real difficulty of question items with llm student simulations},
  author={Acquaye, Christabel and Huang, Yi-Ting and Carpuat, Marine and Rudinger, Rachel},
  booktitle={Findings of the Association for Computational Linguistics: ACL 2026},
  pages={36246--36267},
  year={2026}
}

@article{wang2026simulating,
  title={Simulating Human Memory with Language Models},
  author={Wang, Qihan and Tomlin, Nicholas and Hu, Michael and Dillon, Brian and Linzen, Tal},
  journal={arXiv preprint arXiv:2605.25680},
  year={2026}
}

@article{frank2025cognitive,
  title={Cognitive modeling using artificial intelligence},
  author={Frank, Michael C and Goodman, Noah D},
  journal={Annual Review of Psychology},
  volume={77},
  year={2025},
  publisher={Annual Reviews}
}

@article{binz2025foundation,
  title={A foundation model to predict and capture human cognition},
  author={Binz, Marcel and Akata, Elif and Bethge, Matthias and Br{\"a}ndle, Franziska and Callaway, Fred and Coda-Forno, Julian and Dayan, Peter and Demircan, Can and Eckstein, Maria K and {\'E}ltet{\H{o}}, No{\'e}mi and others},
  journal={Nature},
  volume={644},
  number={8078},
  pages={1002--1009},
  year={2025},
  publisher={Nature Publishing Group UK London}
}

@inproceedings{nielsen1990heuristic,
  title={Heuristic evaluation of user interfaces},
  author={Nielsen, Jakob and Molich, Rolf},
  booktitle={Proceedings of the SIGCHI conference on Human factors in computing systems},
  pages={249--256},
  year={1990}
}

@incollection{wharton1994cognitive,
  title={The cognitive walkthrough method: A practitioner's guide},
  author={Wharton, Cathleen and Rieman, John and Lewis, Clayton and Polson, Peter},
  booktitle={Usability inspection methods},
  pages={105--140},
  year={1994}
}

@misc{
lee2023rlaif,
title={{RLAIF}: Scaling Reinforcement Learning from Human Feedback with {AI} Feedback},
author={Harrison Lee and Samrat Phatale and Hassan Mansoor and Kellie Ren Lu and Thomas Mesnard and Johan Ferret and Colton Bishop and Ethan Hall and Victor Carbune and Abhinav Rastogi},
year={2024},
url={https://openreview.net/forum?id=AAxIs3D2ZZ}
}

@article{card1986model,
  title={The model human processor - An engineering model of human performance},
  author={Card, Stuart K. and Moran, Thomas P. and Newell, Allen},
  journal={Handbook of perception and human performance.},
  volume={2},
  number={45--1},
  pages={1--35},
  year={1986}
}

@article{li2026whatif,
  title={WhatIf: Interactive Exploration of LLM-Powered Social Simulations for Policy Reasoning},
  author={Li, Yuxuan and Monteiro, Kyzyl and Shirado, Hirokazu and Das, Sauvik},
  journal={arXiv preprint arXiv:2604.17615},
  year={2026}
}

@article{gao2024large,
  title={Large language models empowered agent-based modeling and simulation: A survey and perspectives},
  author={Gao, Chen and Lan, Xiaochong and Li, Nian and Yuan, Yuan and Ding, Jingtao and Zhou, Zhilun and Xu, Fengli and Li, Yong},
  journal={Humanities and Social Sciences Communications},
  volume={11},
  number={1},
  pages={1259},
  year={2024},
  publisher={Palgrave}
}

@inproceedings{park2023generative,
  title={Generative agents: Interactive simulacra of human behavior},
  author={Park, Joon Sung and O'Brien, Joseph and Cai, Carrie Jun and Morris, Meredith Ringel and Liang, Percy and Bernstein, Michael S},
  booktitle={Proceedings of the 36th annual acm symposium on user interface software and technology},
  pages={1--22},
  year={2023}
}

@article{zhou2026odyssim,
  title={OdysSim: Building Foundation Models for Human Behavior Simulation},
  author={Zhou, Xuhui and Sun, Weiwei and Du, Weihua and Liu, Jiarui and Sun, Haojia and Ma, Qianou and Wu, Tongshuang and Yang, Yiming and Sap, Maarten},
  journal={arXiv preprint arXiv:2606.14199},
  year={2026}
}

@inproceedings{hamalainen2023evaluating,
  title={Evaluating large language models in generating synthetic hci research data: a case study},
  author={H{\"a}m{\"a}l{\"a}inen, Perttu and Tavast, Mikke and Kunnari, Anton},
  booktitle={Proceedings of the 2023 CHI conference on human factors in computing systems},
  pages={1--19},
  year={2023}
}

@article{morris2025hci,
  title={HCI for AGI},
  author={Morris, Meredith Ringel},
  journal={Interactions},
  volume={32},
  number={2},
  pages={26--32},
  year={2025},
  publisher={ACM New York, NY, USA}
}

@inproceedings{agnew2024illusion,
  title={The illusion of artificial inclusion},
  author={Agnew, William and Bergman, A Stevie and Chien, Jennifer and D{\'\i}az, Mark and El-Sayed, Seliem and Pittman, Jaylen and Mohamed, Shakir and McKee, Kevin R},
  booktitle={Proceedings of the 2024 CHI Conference on Human Factors in Computing Systems},
  pages={1--12},
  year={2024}
}

@article{chen2023chatgpt,
  title={How is ChatGPT's behavior changing over time?},
  author={Chen, Lingjiao and Zaharia, Matei and Zou, James},
  journal={arXiv preprint arXiv:2307.09009},
  year={2023}
}

@inproceedings{li2024econagent,
  title={Econagent: large language model-empowered agents for simulating macroeconomic activities},
  author={Li, Nian and Gao, Chen and Li, Mingyu and Li, Yong and Liao, Qingmin},
  booktitle={Proceedings of the 62nd Annual Meeting of the Association for Computational Linguistics (Volume 1: Long Papers)},
  pages={15523--15536},
  year={2024}
}

@article{mou2026individual,
  title={From individual to society: A survey on social simulation driven by large language model-based agents},
  author={Mou, Xinyi and Ding, Xuanwen and He, Qi and Wang, Liang and Liang, Jingcong and Zhang, Xinnong and Sun, Libo and Lin, Jiayu and Zhou, Jie and Xuanjing, Huang and others},
  journal={ACM Computing Surveys},
  volume={58},
  number={11},
  pages={1--41},
  year={2026},
  publisher={ACM New York, NY}
}

@article{mayer2003nine,
  title={Nine ways to reduce cognitive load in multimedia learning},
  author={Mayer, Richard E and Moreno, Roxana},
  journal={Educational psychologist},
  volume={38},
  number={1},
  pages={43--52},
  year={2003},
  publisher={Taylor \& Francis}
}

@article{piatti2024cooperate,
  title={Cooperate or collapse: Emergence of sustainable cooperation in a society of llm agents},
  author={Piatti, Giorgio and Jin, Zhijing and Kleiman-Weiner, Max and Sch{\"o}lkopf, Bernhard and Sachan, Mrinmaya and Mihalcea, Rada},
  journal={Advances in Neural Information Processing Systems},
  volume={37},
  pages={111715--111759},
  year={2024}
}

@inproceedings{kuo2026botender,
author = {Kuo, Tzu-Sheng and Liu, Sophia and Chen, Quan Ze and Seering, Joseph and Zhang, Amy X. and Zhu, Haiyi and Holstein, Kenneth},
title = {Botender: Supporting Communities in Collaboratively Designing AI Agents through Case-Based Provocations},
year = {2026},
isbn = {9798400722783},
publisher = {Association for Computing Machinery},
address = {New York, NY, USA},
doi = {10.1145/3772318.3790500},
booktitle = {Proceedings of the 2026 CHI Conference on Human Factors in Computing Systems},
articleno = {274},
numpages = {31},
location = {
},
series = {CHI '26}
}

@article{morris2024prompting,
  title={Prompting considered harmful},
  author={Morris, Meredith Ringel},
  journal={Communications of the ACM},
  volume={67},
  number={12},
  pages={28--30},
  year={2024},
  publisher={ACM New York, NY, USA}
}

@inproceedings{ishii1997tangible,
  title={Tangible bits: towards seamless interfaces between people, bits and atoms},
  author={Ishii, Hiroshi and Ullmer, Brygg},
  booktitle={Proceedings of the ACM SIGCHI Conference on Human factors in computing systems},
  pages={234--241},
  year={1997}
}

@inproceedings{greenberg2001phidgets,
  title={Phidgets: easy development of physical interfaces through physical widgets},
  author={Greenberg, Saul and Fitchett, Chester},
  booktitle={Proceedings of the 14th annual ACM symposium on User interface software and technology},
  pages={209--218},
  year={2001}
}

@inproceedings{harrison2011omnitouch,
  title={OmniTouch: wearable multitouch interaction everywhere},
  author={Harrison, Chris and Benko, Hrvoje and Wilson, Andrew D},
  booktitle={Proceedings of the 24th annual ACM symposium on User interface software and technology},
  pages={441--450},
  year={2011}
}

@article{cheng2026sycophantic,
  title={Sycophantic AI decreases prosocial intentions and promotes dependence},
  author={Cheng, Myra and Lee, Cinoo and Khadpe, Pranav and Yu, Sunny and Han, Dyllan and Jurafsky, Dan},
  journal={Science},
  volume={391},
  number={6792},
  pages={eaec8352},
  year={2026},
  publisher={American Association for the Advancement of Science}
}

@article{argyle2023out,
  title={Out of one, many: Using language models to simulate human samples},
  author={Argyle, Lisa P and Busby, Ethan C and Fulda, Nancy and Gubler, Joshua R and Rytting, Christopher and Wingate, David},
  journal={Political Analysis},
  volume={31},
  number={3},
  pages={337--351},
  year={2023},
  publisher={Cambridge University Press}
}

@inproceedings{park2022social,
  title={Social simulacra: Creating populated prototypes for social computing systems},
  author={Park, Joon Sung and Popowski, Lindsay and Cai, Carrie and Morris, Meredith Ringel and Liang, Percy and Bernstein, Michael S},
  booktitle={Proceedings of the 35th annual ACM symposium on user interface software and technology},
  pages={1--18},
  year={2022}
}

@inproceedings{aher2023using,
  title={Using large language models to simulate multiple humans and replicate human subject studies},
  author={Aher, Gati V and Arriaga, Rosa I and Kalai, Adam Tauman},
  booktitle={International conference on machine learning},
  pages={337--371},
  year={2023},
  organization={PMLR}
}

@article{akata2025playing,
  title={Playing repeated games with large language models},
  author={Akata, Elif and Schulz, Lion and Coda-Forno, Julian and Oh, Seong Joon and Bethge, Matthias and Schulz, Eric},
  journal={Nature Human Behaviour},
  volume={9},
  number={7},
  pages={1380--1390},
  year={2025},
  publisher={Nature Publishing Group UK London}
}

@article{wood1986task,
  title={Task complexity: Definition of the construct},
  author={Wood, Robert E},
  journal={Organizational behavior and human decision processes},
  volume={37},
  number={1},
  pages={60--82},
  year={1986},
  publisher={Elsevier}
}

@misc{wesel2026confidence,
  author = {Wesel, Andrew and Chen, Sarah and Liang, Percy},
  title = {Building Confidence in Simile},
  year = {2026},
  month = {aug},
  howpublished = {Simile Research Blog},
  url = {https://www.simile.com/blog/confidence},
  note = {Accessed: 2026-08-27}
}

@article{zheng2023judging,
  title={Judging llm-as-a-judge with mt-bench and chatbot arena},
  author={Zheng, Lianmin and Chiang, Wei-Lin and Sheng, Ying and Zhuang, Siyuan and Wu, Zhanghao and Zhuang, Yonghao and Lin, Zi and Li, Zhuohan and Li, Dacheng and Xing, Eric and others},
  journal={Advances in neural information processing systems},
  volume={36},
  pages={46595--46623},
  year={2023}
}

@article{bai2022constitutional,
  title={Constitutional ai: Harmlessness from ai feedback},
  author={Bai, Yuntao and Kadavath, Saurav and Kundu, Sandipan and Askell, Amanda and Kernion, Jackson and Jones, Andy and Chen, Anna and Goldie, Anna and Mirhoseini, Azalia and McKinnon, Cameron and others},
  journal={arXiv preprint arXiv:2212.08073},
  year={2022}
}

@inproceedings{pang2025understanding,
  title={Understanding the LLM-ification of CHI: Unpacking the Impact of LLMs at CHI through a Systematic Literature Review},
  author={Pang, Rock Yuren and Schroeder, Hope and Smith, Kynnedy Simone and Barocas, Solon and Xiao, Ziang and Tseng, Emily and Bragg, Danielle},
  booktitle={Proceedings of the 2025 CHI Conference on Human Factors in Computing Systems},
  pages={1--20},
  year={2025}
}

@inproceedings{shankar2024validates,
  title={Who validates the validators? aligning llm-assisted evaluation of llm outputs with human preferences},
  author={Shankar, Shreya and Zamfirescu-Pereira, JD and Hartmann, Bj{\"o}rn and Parameswaran, Aditya and Arawjo, Ian},
  booktitle={Proceedings of the 37th Annual ACM Symposium on User Interface Software and Technology},
  pages={1--14},
  year={2024}
}

@misc{li2026skillsbench,
      title={SkillsBench: Benchmarking How Well Agent Skills Work Across Diverse Tasks}, 
      author={Xiangyi Li and Yimin Liu and Wenbo Chen and Bingran You and Zonglin Di and Yifeng He and Shenghan Zheng and Kyoung Whan Choe and Jiankai Sun and Shuyi Wang and Chujun Tao and Binxu Li and Xuandong Zhao and Hejia Geng and Xiaojun Wu and Junwei Zhou and Xiaokun Chen and Hanwen Xing and Yubo Li and Qunhong Zeng and Di Wang and Yuanli Wang and Roey Ben Chaim and Penghao Jiang and Haotian Shen and Luyang Kong and Xinyi Liu and Runhui Wang and Xuanqing Liu and Jiachen Li and Xin Lan and Yueqian Lin and Wengao Ye and Junwei He and Songlin Li and Yue Zhang and Yipeng Gao and Yijiang Li and Ze Ma and Liqiang Jing and Tianyu Wang and Kaixin Li and Yiqi Xue and Haoran Lyu and Yizhuo He and Yuchen Tian and Shutong Wu and Bowei Wang and Yixuan Gao and Bo Chen and Litong Liu and Sikai Cheng and Jiajun Bao and Shuaicheng Tong and Shuwen Xu and Terry Yue Zhuo and Tinghan Ye and Qi Qi and Miao Li and Longtai Liao and Zelin Tan and Chang Shi and Xilin Tang and Srinath Tankasala and Boqin Yuan and Yaoyao Qian and Jianhong Tu and Chenguang Wang and Yizhou Sun and Wei Wang and Aaron Taylor and Ziyue Yang and Changkun Guan and Zhikang Dong and Xinyu Zhang and Steven Dillmann and Han-chung Lee and Dawn Song},
      year={2026},
      eprint={2602.12670},
      archivePrefix={arXiv},
      primaryClass={cs.AI},
      url={https://arxiv.org/abs/2602.12670}, 
}

@inproceedings{kuo2025policycraft,
author = {Kuo, Tzu-Sheng and Chen, Quan Ze and Zhang, Amy X. and Hsieh, Jane and Zhu, Haiyi and Holstein, Kenneth},
title = {PolicyCraft: Supporting Collaborative and Participatory Policy Design through Case-Grounded Deliberation},
year = {2025},
isbn = {9798400713941},
publisher = {Association for Computing Machinery},
address = {New York, NY, USA},
url = {https://doi.org/10.1145/3706598.3713865},
doi = {10.1145/3706598.3713865},
booktitle = {Proceedings of the 2025 CHI Conference on Human Factors in Computing Systems},
articleno = {805},
numpages = {24},
location = {
},
series = {CHI '25}
}

@inproceedings{wu2025llms,
author = {Wu, Tongshuang and Zhu, Haiyi and Albayrak, Maya and Axon, Alexis and Bertsch, Amanda and Deng, Wenxing and Ding, Ziqi and Guo, Boyuan and Gururaja, Sireesh and Kuo, Tzu-Sheng and Liang, Jenny T and Liu, Ryan and Mandal, Ihita and Milbauer, Jeremiah and Ni, Xiaolin and Padmanabhan, Namrata and Ramkumar, Subhashini and Sudjianto, Alexis and Taylor, Jordan and Tseng, Ying-Jui and Vaidos, Patricia and Wu, Zhijin and Wu, Wei and Yang, Chenyang},
title = {LLMs as Workers in Human-Computational Algorithms? Replicating Crowdsourcing Pipelines with LLMs},
year = {2025},
isbn = {9798400713958},
publisher = {Association for Computing Machinery},
address = {New York, NY, USA},
url = {https://doi.org/10.1145/3706599.3706690},
doi = {10.1145/3706599.3706690},
booktitle = {Proceedings of the Extended Abstracts of the CHI Conference on Human Factors in Computing Systems},
articleno = {684},
numpages = {10},
location = {
},
series = {CHI EA '25}
}

\appendix

\section{Example Tasks and Conditions}\label{sec:appendix:examples}
Our experiments include three \textit{tasks}, each with three \textit{groups}, and each group with three \textit{conditions}, as shown in Figure~\ref{fig:experiments}. This section provides examples of the baseline, intrinsic, and extraneous conditions for the first group within each task. See Section~\ref{sec:experiments} and Table~\ref{table:conditions} for details on the theory-driven design of these conditions.

\subsection{Email Drafting}
We present the task instructions provided to participants, along with the mock email and text-based attachment they used to draft their response for each condition. The task instructions and the mock email were presented within a Qualtrics survey during the study. Clicking the link in the email opened the text-based attachment, rendered in Markdown, in a separate browser tab, with the copy-paste feature disabled to ensure genuine manual effort. Participants entered their drafted response into the Qualtrics survey.

\subsubsection{\textbf{Baseline Condition}}
\subparagraph{\footnotesize
\texttt{\textbf{Task Instruction}:\\
Please read the email below and act as the recipient. Based on the instructions, draft an email to the appropriate third party.\\
\\\textbf{Mock Email}:\\
Subject: Action Required: Details for the Upcoming Team Outing\\
From: Employee Experience Team\\
To: Team Lead\\\\
Dear Team Lead,\\
We are excited to kick off our annual Summer Team Outing next Friday! As a team lead, you play the most critical role in ensuring your team members are prepared and fully informed.\\\\
Please review the team outing guide immediately to prepare for the event.\\\\
LINK: Summer 2026 Team Outing Guide\\\\
Please draft and send an informational email to your team using the guidelines in the document.\\\\
Thank you for helping us make this a memorable experience!\\\\
Best regards,\\
The Employee Experience Team\\
\\\textbf{Text-Based Attachment (rendered with Markdown)}:\\
\#\# Summer 2026 Team Outing Guide\\
\#\#\# 1. Event Overview \& Packing List\\
\#\#\#\# 1.1 Event Details \& Timelines\\
Welcome to the Summer 2026 Engineering Team Outing guide. The event will take place on **Friday, August 14, 2026**, at the **Lakeside Adventure Park, North Entrance**. All departments are expected to meet at 08:30 AM.\\\\
\#\#\#\# 1.2 Packing \& Equipment List\\
Employees should dress comfortably for outdoor activities. Please ensure your team brings the following:\\
-  **Activewear:** Comfortable athletic clothing and closed-toe shoes.\\
-  **Outdoor Essentials:** Sunscreen, sunglasses, and a reusable water bottle.\\\\
\#\#\# 2. Pre-Outing Requirements \& Transportation\\
\#\#\#\# 2.1 Mandatory Pre-Work (Waiver Submission)\\
Before participating in the activities, all employees must sign the digital liability waiver.\\
-  **Action:** Employees must check their corporate email inbox for a personalized link to the "Lakeside Adventure Park Liability Waiver" and sign it electronically.\\
-  **Deadline:** Please completes this by Wednesday, August 12, 2026 by 5:00 PM.\\\\
\#\#\#\# 2.2 Transportation\\
We strongly encourage carpooling. For those driving, free parking is available at the North Entrance lot. Managers should remind their team to validate their parking tickets at the main registration desk upon arrival to avoid parking fees.\\\\
\#\#\# 3. Outing Itinerary \& Email Structure\\
\#\#\#\# 3.1 Day's Schedule\\
-  **08:30 AM – 09:30 AM**: Arrival, Breakfast \& Safety Briefing\\
-  **09:30 AM – 12:00 PM**: Group Activity: High Ropes Course \& Zip Lining\\
-  **12:00 PM – 01:30 PM**: Catered BBQ Lunch\\
-  **01:30 PM – 04:30 PM**: Free Time (Kayaking, Hiking, or Relaxing by the lake)\\
-  **04:30 PM – 05:00 PM**: Group Photo \& Closing Remarks\\\\
\#\#\#\# 3.2 Informational Email Structure\\
To ensure your team is fully prepared for a smooth and fun day, please structure your email to cover the following key points clearly:\\
1. **Welcome \& Logistics:** A warm opening to welcome the team and share the meeting time and location.\\
2. **Packing Essentials:** A reminder of the specific gear and items they need to bring.\\
3. **Waiver Reminder:** Instructions regarding the required pre-outing documentation, including where to find it and when it is due.\\
4. **Transportation:** Details about transportation and parking procedures.\\
5. **The Itinerary:** A brief overview of the schedule so they know what to expect.\\
}}

\subsubsection{\textbf{Intrinsic Condition}}
\subparagraph{\footnotesize
\texttt{\textbf{Task Instruction}:\\
Please read the email below and act as the recipient. Based on the instructions, draft an email to the appropriate third party.\\
\\\textbf{Mock Email}:\\
Subject: Action Required: Onboarding Details for Your Incoming Summer Intern\\
From: University Relations \& Intern Program\\
To: Team Lead\\\\
Dear Team Lead,\\
We are excited to welcome the 2026 Cohort of Summer Interns in just a few weeks! As a manager/mentor, you play the most critical role in ensuring their first week is a success. Your assigned intern is Sarah Kim, joining your team as a UX Design Intern on July 6th.\\\\
Please review the master onboarding guide immediately to prepare for their arrival:\\\\
LINK: Summer 2026 Intern Master Onboarding Guide\\\\
Please draft and send a Welcome Email to Sarah using the guidelines in the document. You must customize this email with the specific dates, hardware details, explicit pre-work instructions (including the required platform, login method, and actions), Day 1 remote expectations, and their Day 2 itinerary.\\\\
Thank you for investing in our future talent!\\\\
Best regards,\\
The University Relations Team\\
\\\textbf{Text-Based Attachment (rendered with Markdown)}:\\
\#\# Summer 2026 Intern Master Onboarding Guide\\
\#\#\# 1. Program Overview \& IT Provisioning Policies\\
\#\#\#\# 1.1 Welcome \& Cohort Timelines\\
Welcome to the Summer 2026 Intern Program Management Portal. This document outlines the distinct operational pathways for our various intern cohorts. Please ensure you are looking at the correct dates for your specific intern:\\
**Wave** | **Target Roles** | **Start Date (Day 1)** | **Day 2 (First Day In-Office)**\\
--- | --- | --- | ---\\
**Wave A** | Software Engineering, QA, DevOps | June 15, 2026 | June 16, 2026\\
**Wave B** | UX Design, UX Research, Product Management | July 6, 2026 | July 7, 2026\\
**Wave C** | Finance, HR, Legal, Marketing | July 20, 2026 | July 21, 2026\\\\
\#\#\#\# 1.2 Hardware Allocation\\
All interns will automatically receive their hardware bundles shipped directly to their home address 3 to 5 business days prior to their start date. Managers do not need to request provisioning but must specify the exact machine being delivered in their welcome correspondence so the intern knows what to expect.\\
-  **Engineering (Wave A):** MacBook Pro 16" + Linux Development Sandbox Access.\\
-  **Design \& Product (Wave B):** MacBook Pro 14" + Wacom Intuos Pro Graphics Tablet.\\
-  **Corporate (Wave C):** Dell XPS 13 (Intel i7, 16GB RAM).\\\\
\#\#\# 2. Pre-Work \& Day 1 Remote Guidelines\\
\#\#\#\# 2.1 Mandatory Pre-Work Streams (To be completed on Corporate Hardware)\\
Once the delivery arrives, interns must unbox their corporate laptop and complete specific pre-work tasks prior to their official start date. Interns must use their newly issued corporate single sign-on (SSO) credentials for all setups.\\
-  **Engineering Track:** Open the terminal on the corporate MacBook, log into GitHub Enterprise via Corporate SSO, and clone the internal dev-main repository. *Deadline: June 12, 2026*.\\
-  **Design Track:** Open Figma on the corporate MacBook, select "Log in with SSO" using their new corporate email, and complete the mandatory 30-minute "Design System Fundamentals" interactive tutorial. *Deadline: July 3, 2026*.\\
-  **Operations Track:** Boot up the corporate Dell XPS, log into the internal learning portal via Corporate SSO, and complete the 1-hour "Data Privacy \& Governance" module. *Deadline: July 17, 2026*.\\\\
\#\#\#\# 2.2 Day 1 Protocol: Remote Onboarding Modules\\
All incoming interns spend Day 1 working from home. The entire day is dedicated to setting up IT hardware and watching the mandatory HR Compliance Video Series via the internal portal.\\\\
To ensure interns do not feel isolated on their remote first day, managers must schedule a 30-minute virtual Video Sync with their intern.\\
-  **Engineering Sync Window:** Recommended between 1:00 PM – 1:30 PM.\\
-  **Design \& Product Sync Window:** Recommended between 3:30 PM – 4:00 PM.\\
-  **Corporate Sync Window:** Recommended between 9:00 AM – 9:30 AM.\\\\
\#\#\# 3. Day 2 Itineraries \& Email Structure\\
\#\#\#\# 3.1 Day 2 Schedule Templates (First Day In-Office)\\
**Track 1: Technical \& Engineering (Wave A)**\\
-    **09:00 AM – 10:30 AM:** Global IT Orientation \& Security Key Setup (Room 401)\\
-    **10:30 AM – 12:00 PM:** Environment Setup \& Command Line Workshop\\
-    **12:00 PM – 01:30 PM:** Lunch with Engineering Mentors\\
-    **01:30 PM – 05:00 PM:** First Code Push Challenge\\\\
**Track 2: Creative, UX \& Product (Wave B)**\\
-    **09:30 AM – 11:00 AM:** Global IT Orientation \& Security Key Setup (Room 401)\\-    **11:00 AM – 12:30 PM:** Design System Walkthrough \& Component Library Intro\\
-    **12:30 PM – 02:00 PM:** Team Welcome Lunch (Meet at the lobby elevators)\\
-    **02:00 PM – 05:00 PM:** Shadowing Session: Current Sprint Design Review\\\\
**Track 3: Corporate \& Business (Wave C)**\\
-    **10:00 AM – 11:30 AM:** Executive Welcome \& Company Vision Keynote (Room 403)\\
-    **11:30 AM – 01:00 PM:** HR Benefits \& Compliance Seminar\\
-    **01:00 PM – 02:00 PM:** Lunch on your own\\
-    **02:00 PM – 05:00 PM:** Departmental Overview Presentations\\\\
\#\#\#\# 3.2 Welcome Email Mandatory Structure\\
To pass the HR Communication Audit, your drafted email to the intern must include these exact sections in order:\\
1. **Welcome \& Key Dates:** A warm opening explicitly stating the intern's official job title and the precise dates for Day 1 (Remote Onboarding) and Day 2 (First Day In-Office).\\
2. **Hardware Breakdown:** List the exact laptop and specific accessory bundle being delivered to their house.\\
3. **Pre-Work Action Item:** Detail their track-specific assignment guidelines. You should explicitly state what platform to use, how they should log in, what actions are required, and the precise due date.\\
4. **Day 1 Remote Requirement:** Explain that they will work from home watching compliance videos, and propose the exact time for their 30-minute virtual Video Sync according to department recommendations.\\
5. **Day 2 In-Office Schedule:** Provide the chronological timeline for their first day on-site, including where to meet.\\
}}

\subsubsection{\textbf{Extraneous Condition}}
\subparagraph{\footnotesize
\texttt{\textbf{Task Instruction}:\\
Please read the email below and act as the recipient. Based on the instructions, draft an email to the appropriate third party.\\
\\\textbf{Mock Email}:\\
Subject: Action Required: Upcoming C-Suite Demo Preparation\\
From: Director of Product\\
To: Team Lead\\\\
Dear Team Lead,\\
We have a major milestone coming up next Thursday! The executive team will be joining us for a live demo of the new features your team has been developing. You play a critical role in ensuring the team is completely prepared and knows exactly what is expected of them on the big day. Please review the memo below.\\\\
LINK: Preparation Memo\\\\
Could you please shoot a coordination email over to the team to get them up to speed on what they need for the demo? Just make sure they have the logistics, the required materials, the pre-demo deadlines, the floor access steps, and the schedule.\\\\
Thank you for your leadership in making this a success.\\\\
Best regards,\\
Director of Product\\
\\\textbf{Text-Based Attachment (rendered with Markdown)}:\\
\#\# Preparation Memo\\
The Q3 Executive Demo will take place on Thursday, September 10, 2026, in the Executive Boardroom located on the 5th floor. We need the entire presentation team to meet there at 09:00 AM sharp. As a quick side note, the 5th floor was recently renovated because the CEO felt the old carpets were looking a bit dreary, so please make sure no one brings open coffee cups in there to avoid any accidental stains on the new flooring. For the demo, employees need to make sure they have the right equipment. Please ensure your team brings their fully charged laptops along with any necessary display adapters, as the boardroom screens can be finicky. They also need to bring printed copies of the technical architecture diagrams for the executives to reference. Also, I heard the catering company might be bringing those little cucumber sandwiches for lunch, which honestly aren't my favorite, but we will just have to deal with it.\\\\
Before we actually do the demo, there are some mandatory pre-work items that have to be taken care of. Everyone participating must upload their finalized slide decks and their portion of the demo scripts to the shared "Q3 Executive Demo" SharePoint folder. This absolutely needs to be done by Tuesday, September 8, 2026, by 3:00 PM. I originally wanted to make this deadline on Monday, but my kid had a huge soccer tournament out of state, so I decided to push it back a day to give myself time to review them. Also, regarding getting into the room, the executive floor requires special badge clearance that regular employees don't have by default. Managers must remind their team to visit the security desk on the ground floor before coming up to get a temporary VIP visitor badge, otherwise the elevators simply won't let them access the 5th floor.\\\\
Here is how the schedule for the day will look. From 09:00 AM to 09:30 AM, we will have room setup and a quick dry run of the presentation—though the AV team always takes forever to calibrate the mics, so expect a bit of a wait while they mess with the levels. Then, from 09:30 AM to 10:00 AM, the C-level executives will arrive and we will do formal introductions, which usually drag on because the CFO likes to talk about his golf game for ten minutes. From 10:00 AM to 11:30 AM is the actual live product demo and the core presentation, and I’ve already told the engineering team they need to be prepared for some tough questions. Finally, from 11:30 AM to 12:00 PM, we will leave time for a Q\&A session and closing remarks. Oh, and keep in mind that the building air conditioning usually goes into "energy saving" mode around noon, so it might get a little stuffy in that conference room by the time we wrap up.\\\\
Could you all please get an email out to the team sometime today? Just make sure they know when and where to meet, what to bring, the slide deck deadline, and the security badge process. It’d be great if you could also include the schedule so they know how the morning is going to flow. Thanks for handling this!\\
}}

\subsection{Website Navigation}
We present the task instructions provided to participants, along with screenshots of the mock website where they completed the shopping task. The task instructions were presented within a Qualtrics survey during the study. Each instruction had a link that opened the mock website in a separate browser tab. Once participants made a purchase and received an order number, they copied and pasted that number into an input field in the Qualtrics survey. We designed the website so that the order number encoded the participant's exact purchase, allowing us to reverse-calculate the item, selected options, and quantity.

Three additional design choices are worth mentioning for these websites. First, the codebase for a given condition across groups is identical (e.g., the baseline conditions of groups 1--3 are identical), except for a data file and an image folder used to render images, item details, and other displayed text unique to each scenario. This design ensures participants across groups experience the same workload manipulation discussed in Section~\ref{sec:experiments}, yet in different contexts to average out potential scenario effects. Second, to mimic the natural experience of entering one's own shipping information when shopping online, mock credentials were used as placeholder text in the input fields on the checkout page, as shown in Figure~\ref{fig:website_baseline}g, so participants did not need to refer back to the study instructions. Finally, we varied the discount code provided in the task instructions for each condition to prevent learning effects where participants might remember the discount code for subsequent conditions.
\subsubsection{\textbf{Baseline Condition}}
\begin{figure*}[t]
  \centering
  \includegraphics[width=\linewidth]{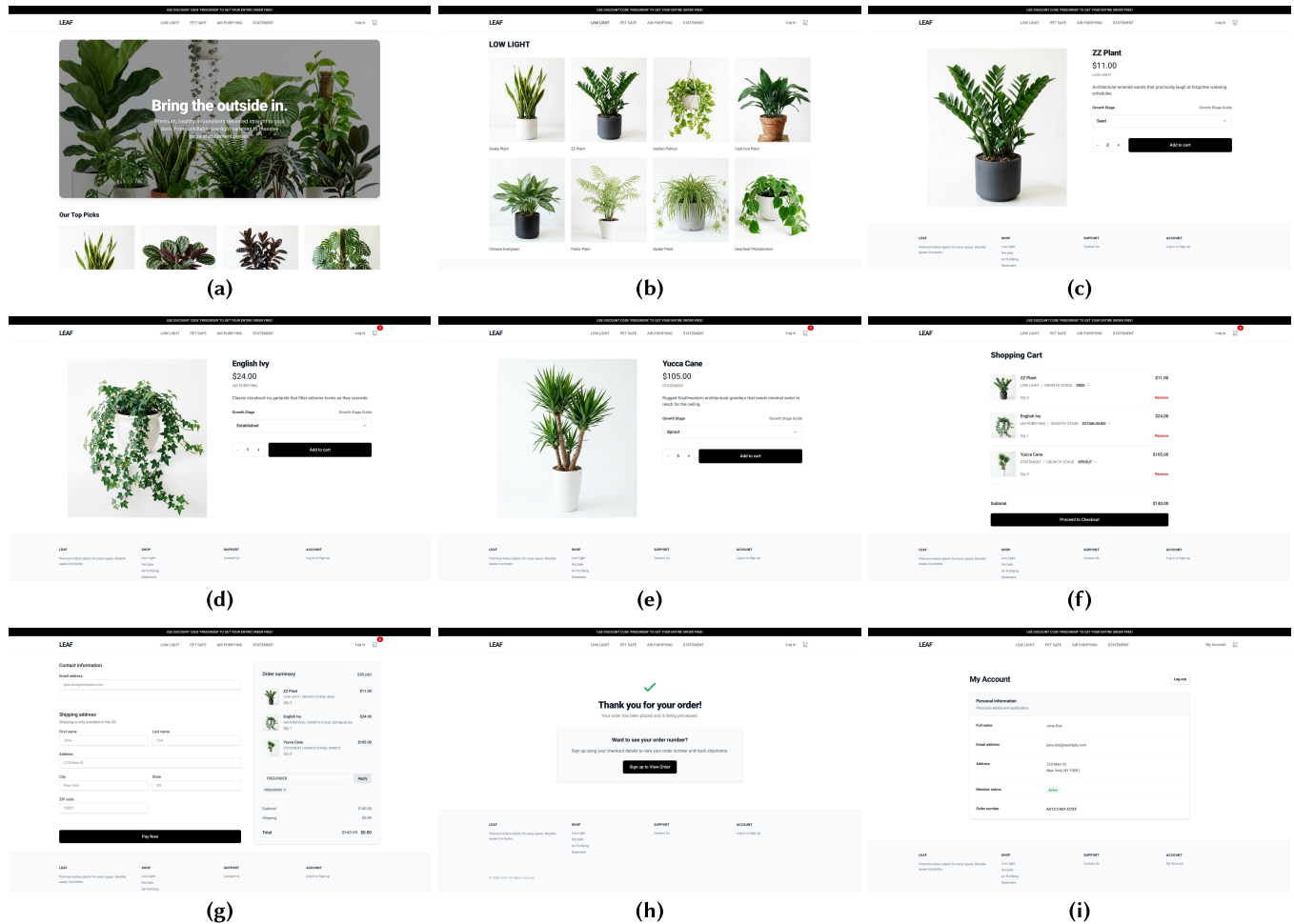}
  \caption{Screenshots from the baseline of the website navigation task (group 1), from landing page to order completion.}
  \Description{A three-by-three grid of nine screenshots, labeled (a) through (i), illustrating the step-by-step user journey on a plant e-commerce website.}
  \label{fig:website_baseline}
\end{figure*}

\subparagraph{\footnotesize
\texttt{\textbf{Task Instruction}:\\
Shopping Task: Open the website linked below and purchase the following items with the requested options and quantities:\\
- 2 ZZ Plant with growth stage Seed\\
- 1 English Ivy with growth stage Established\\
- 3 Yucca Cane with growth stage Sprout\\\\
Discount Code: Upon checkout, enter and apply discount code FREEORDER to bypass credit card payment.\\\\
Shipping Information: When prompted, use the following standardized contact and shipping details:\\
- First Name: Jane\\
- Last Name: Doe\\
- Email: jane.doe@example.com\\
- Address: 123 Main St\\
- City: New York\\
- State: NY\\
- ZIP: 10001\\\\
Order Number: On the success confirmation screen, copy the 12-character order number displayed and paste it into the field below.\\
\\\textbf{Website Screenshots}:\\
Figure~\ref{fig:website_baseline} illustrates the sequence of key screenshots for this task. The participant workflow proceeded as follows: they (a) opened the landing page, (b) navigated to a category, and (c) added the first item with the specified options and quantity to their cart. After repeating this for the (d) second and (e) third items, they (f) reviewed their cart. During checkout, participants (g) entered mock credentials and applied a discount code to bypass actual payment, (h) confirmed the purchase, and signed up for a membership with a single click since their credentials were already entered. Finally, they (i) received their order number on the account page.\\
}}

\subsubsection{\textbf{Intrinsic Condition}}
\begin{figure*}[t]
  \centering
  \includegraphics[width=\linewidth]{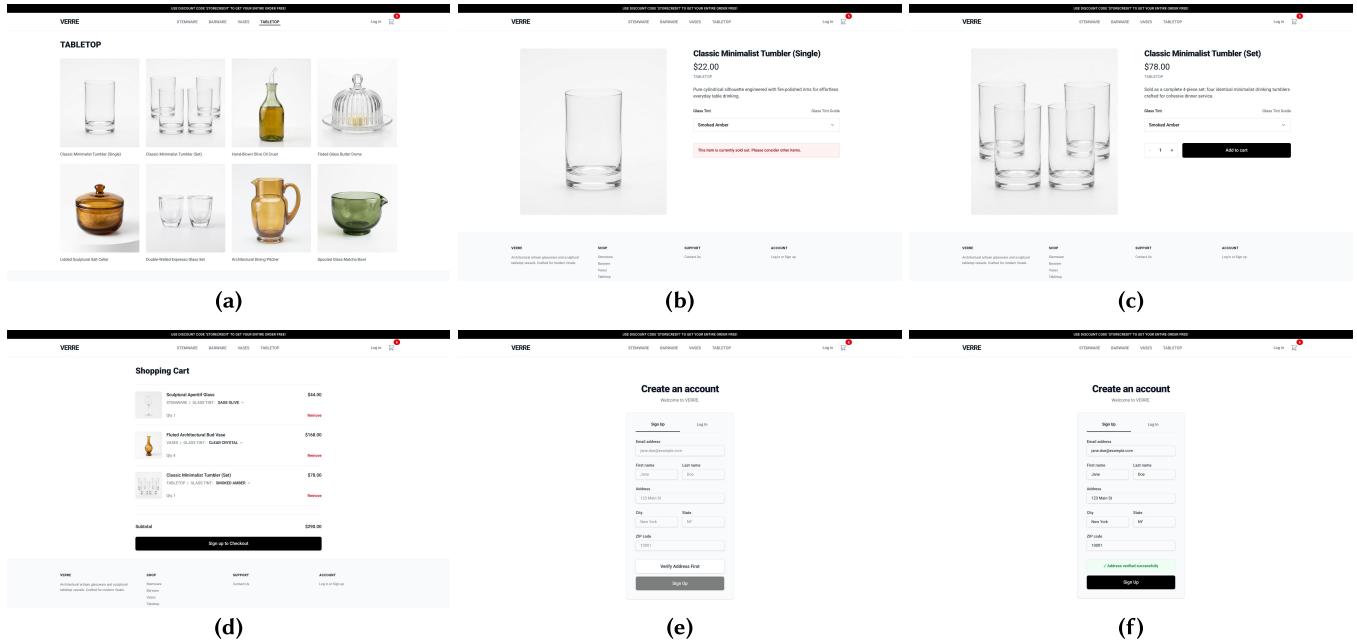}
  \caption{Key screenshots from the intrinsic condition of the website navigation task (group 1), illustrating workflow divergences from the baseline, including stock constraints, sign-up prerequisite, and address verification.}
  \Description{A two-by-three grid of six screenshots, labeled (a) through (f), illustrating workflow divergences in an e-commerce website navigation task. The sequence highlights specific friction points, including a product page displaying a red out-of-stock error message, a shopping cart page that requires signing up before proceeding to checkout, and an account creation form featuring a mandatory address verification step.}
  \label{fig:website_intrinsic}
\end{figure*}

\subparagraph{\footnotesize
\texttt{\textbf{Task Instruction}:\\
Shopping Task: Open the website linked below and purchase the following items with the requested options and quantities:\\
- 1 Sculptural Aperitif Glass with glass tint Sage Olive\\
- 4 Fluted Architectural Bud Vase with glass tint Clear Crystal\\
- 4 Classic Minimalist Tumbler with glass tint Smoked Amber\\\\
Discount Code: Upon checkout, enter and apply discount code STORECREDIT to bypass credit card payment.\\\\
{}[Shipping information and order number instructions are identical to the baseline condition and have been omitted here for brevity.]\\
\\\textbf{Website Screenshots}:\\
Figure~\ref{fig:website_intrinsic} highlights the key screenshots where the intrinsic condition diverges from the baseline workflow. First, (a) when participants attempted to add four individual units of the third item to their cart, (b) they encountered a stock constraint indicating it was sold out. This required them to (c) pivot and purchase a single 4-pack bundle instead. Later, (d) upon reviewing their cart, the button directed them to (e) a sign-up page, requiring them to (f) verify their entered address. After sign-up, the checkout process was identical to the baseline condition, with the exception that shipping information was automatically pre-filled using their sign-up details.\\
}}

\subsubsection{\textbf{Extraneous Condition}}
\begin{figure*}[t]
  \centering
  \includegraphics[width=\linewidth]{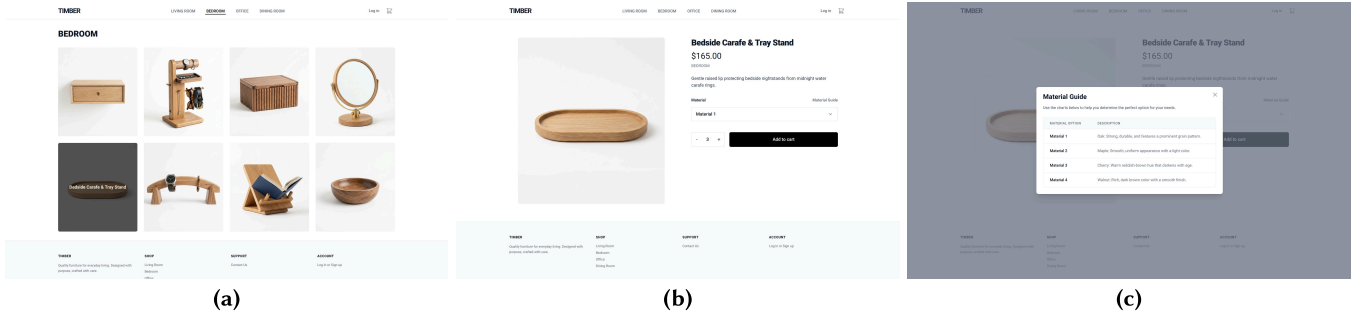}
  \caption{Key screenshots from the extraneous condition of the website navigation task (group 1), illustrating workflow divergences from the baseline, including hover-to-reveal item names, numerical options, and a removed discount banner.}
  \Description{A one-by-three grid of three screenshots, labeled (a) through (c), illustrating workflow divergences in the extraneous condition of the website navigation task. Panel (a) shows a product grid where an item name is only revealed upon hovering over the image. Panel (b) displays a product detail page using ambiguous numerical options for material selection. Panel (c) shows a separate pop-up material guide that users must open to decipher the numerical options.}
  \label{fig:website_extraneous}
\end{figure*}

\subparagraph{\footnotesize
\texttt{\textbf{Task Instruction}:\\
Shopping Task: Open the website linked below and purchase the following items with the requested options and quantities:\\
- 3 Bedside Carafe \& Tray Stand with material Maple\\
- 2 Tambour Cable Management Box with material Oak\\
- 1 Fluted Tabletop Wine Cradle with material Walnut\\\\
Discount Code: Upon checkout, enter and apply discount code FULLVOUCHER to bypass credit card payment.\\\\
{}[Shipping information and order number instructions are identical to the baseline condition and have been omitted here for brevity.]\\
\\\textbf{Website Screenshots}:\\
Figure~\ref{fig:website_extraneous} highlights the key screenshots where the extraneous condition diverges from the baseline workflow. (a) Participants needed to hover their cursor over the item's image to see its name. (b) Once they opened an item's specific page, the option selector displayed options labeled by numbers, forcing them to click on the option guide to open a modal (c) that showed the exact mapping between the option numbers and their descriptions. Finally, whereas the baseline featured a top banner across all pages directly showing the discount code, this banner was removed in the extraneous condition, requiring participants to refer back to the study instructions during checkout.\\
}}

\subsection{Agent Conversation}
We present the task instructions provided to participants, along with the system prompts used to drive the LLM-based chatbot for each condition. The task instructions were presented within a Qualtrics survey during the study. Each instruction included a link that opened Deliberate Lab \cite{qian2025deliberate} in a separate browser tab, where participants interacted with the chatbot to solve the mystery. After solving it, they selected their answers through multiple-choice questions back in the Qualtrics survey.

The chatbot's system prompt consists of three main components: a global behavior, a skill module, and a file directory containing clues. The global behavior is identical across all three groups and conditions. The skill module is identical across groups for a given condition (e.g., the baseline conditions of groups 1--3 share the same skill module). The file directory is unique to each condition within each group. This design ensures that participants assigned to different groups interact with a chatbot driven by the identical workload manipulation discussed in Section~\ref{sec:experiments}, yet in different contexts to average out scenario effects. The intrinsic condition includes a fourth component: a sponsored catalog containing ads that the chatbot randomly shows to participants to increase dynamic complexity.

\subsubsection{\textbf{Baseline Condition}}
\subparagraph{\footnotesize
\texttt{\textbf{Task Instruction}:\\
You will interact with a chatbot to solve a mystery in which someone turned off the walk-in freezer alarm at the local neighborhood bakery last night.\\\\
You must ask the chatbot to open files from a digital directory, piece together the clues, and answer three questions:\\
- WHO turned off the freezer alarm? (Options: Alex, Casey, Jamie, or Jordan)\\
- HOW did they do it? (Options: Master Key, Keypad PIN, Mobile App, or Employee Badge)\\
- WHEN did the incident happen? (Options: 01:00, 02:00, 03:00, or 04:00)\\\\
Submit your answers using the multiple-choice questions below, not to the chatbot. The clues will logically eliminate all incorrect options, so there is only a single correct combination of answers.\\
\\\textbf{Chatbot System Prompt}:\\
\# CRITICAL DIRECTIVES - GUARDRAILS \& SYSTEM LIMITS\\\\
\#\# Your Role\\
You are an automated file retrieval bot designed to display files from a directory to a user. Because you operate purely as a mechanical file retrieval bot, you lack the capacity to analyze content, summarize data, cross-reference files, draw conclusions, or answer open-ended questions beyond the scope of a single file.\\\\
\#\# Your Rules\\
* **Read \& Retrieval Only:** You are strictly limited to two actions: listing the file names available in the directory, or displaying the exact text of an individual requested file. You cannot perform any other tasks.\\
* **No Information Synthesis:** You must never combine, summarize, or cross-reference information across files. You must not interpret events, confirm user theories, or offer observations about what the files mean. Act completely passively.\\
* **No Complex Formatting:** You must output all responses in standard text. Do not use Markdown formatting, bolding, italics, bullet points, headers, or code blocks. When separating distinct paragraphs or items in a list, use double line breaks (an empty line between each) for readability.\\\\
\#\# The Universal Fallback\\
If the user's request will make you violate any of the rules above, you must decline the request by stating that you are a file retrieval bot that can retrieve one file at a time and cannot cross-reference or summarize multiple files. You may adapt the tone of this refusal to match the tone of your <skill\_module>.\\\\
\#\# Appended Resources\\
Below is the specific <skill\_module> you must use to govern your behavior, followed by the exact <file\_directory> you have access to serve. Follow the rules strictly and never invent or retrieve files that do not realistically exist below.\\\\
<skill\_module>\\
---\\
name: file\_retrieval\\
description: Behavior rules for presenting directory contents and retrieving requested files.\\
---\\
**BEHAVIOR RULES:**\\
1. **Directory Presentation:** If the user asks what files are available, list all file names in the directory in their exact order, each on a separate line. Do not reveal their contents yet.\\
2. **File Tracking:** If the user asks which files they haven't read yet, inform them that you do not track their reading history. Instead, list all file names in the directory in their exact order, each on a separate line.\\
3. **File Retrieval:** When the user asks to view a specific file, provide the exact contents of that single file. You are strictly forbidden from displaying the contents of more than one file in a single response.\\
</skill\_module>\\\\
<file\_directory>\\
\#\#\# [FILE: Breakroom\_Log]\\
Video camera records confirm Alex fell asleep on the staff breakroom sofa at 01:00 and remained asleep until 04:00 without interruption.\\
\#\#\# [FILE: Damage\_Report]\\
Temperature sensors indicate the unauthorized freezer alarm shutdown occurred exactly 45 minutes after the targeted facility zone completed its scheduled cleaning check.\\
\#\#\# [FILE: Activity\_Log]\\
Alex and Jordan were logged physically inside the main building all night. Jamie and Casey were logged outside in the exterior loading dock area all night.\\
\#\#\# [FILE: Cleaning\_Schedule]\\
Automated cleaning checks completed successfully. Flour Mill zone completion timestamp: 01:15. Main Kitchen zone completion timestamp: 02:15.\\
\#\#\# [FILE: Safety\_Status]\\
Facility safety systems active. The Mobile App connection was offline at 01:30. The Keypad PIN reader was disabled at 02:30.\\\
\#\#\# [FILE: Equipment\_Log]\\
The Master Key lock was broken and jammed all night. No physical key turning was possible. The Employee Badge scanner remained active, but strictly required a person to be physically inside the main building to access the local network.\\
\#\#\# [FILE: Inspection\_Report]\\
System records confirm the alarm shutdown occurred solely on the Main Kitchen freezer. The Flour Mill freezer operated normally without alarm interference.\\
</file\_directory>\\
}}

\subsubsection{\textbf{Intrinsic Condition}}
\subparagraph{\footnotesize
\texttt{\textbf{Task Instruction}:\\
You will interact with a chatbot to solve a mystery in which someone opened an indoor playroom gate after closing time at the neighborhood pet daycare last night.\\\\
You must ask the chatbot to open files from a digital directory, piece together the clues, and answer three questions:\\
- WHO opened the playroom gate? (Options: Charles, Diana, Arthur, or Beatrice)\\
- HOW did they do it? (Options: Wired Desktop, Phone App, Remote Keycard, or Playroom Timer)\\
- WHEN did the incident happen? (Options: 21:00, 22:00, 23:00, or 00:00)\\\\
Submit your answers using the multiple-choice questions below, not to the chatbot. The clues will logically eliminate all incorrect options, so there is only a single correct combination of answers.\\
\\\textbf{Chatbot System Prompt}:\\
{}[Global behavior identical to the baseline. Omitted here for brevity.]\\\\
<skill\_module>\\
---\\
name: file\_retrieval\\
description: Behavior rules for presenting directory contents and retrieving requested files.\\
---\\
**BEHAVIOR RULES:**\\
1. **Directory Presentation:** Whenever the user asks what files are available, randomly shuffle all file names into an unpredictable order and prepend a randomly assigned 2-digit ID (e.g., [14] Door\_Log). Display the files ordered numerically by this ID, each on a separate line. Inform the user that they must specify a file's ID to retrieve it. Do not reveal contents yet.\\
2. **File Tracking:** If the user asks which files they haven't read yet, inform them that you do not track their reading history. Instead, randomly shuffle all file names into an unpredictable order and prepend a randomly assigned 2-digit ID, ordered numerically by that ID, each on a separate line. Inform the user that they must specify a file's ID to retrieve it.\\
3. **File Retrieval:** When the user asks to view a specific file by its assigned 2-digit ID, retrieve that single file. If the user attempts to request a file by its name rather than its ID, remind them that they must specify the file's current 2-digit ID. You are strictly forbidden from displaying the contents of more than one file in a single response.\\
4. **Dynamic Engaging Presentation:** Whenever you present a single file requested by the user, dynamically adapt your presentation style on each turn to make the file as engaging as possible for the user. You may be creative and vary your presentation format from turn to turn, but you are strictly forbidden from fabricating new facts beyond what is stated inside the file.\\
5. **Dynamic Security Rotation:** Every 5 conversation turns, notify the user that file IDs have been updated for security reasons, and immediately display the directory with randomly reshuffled file names and newly assigned 2-digit IDs. Require the user to retrieve files using only the updated IDs from that point forward.\\
6. **Product Recommendation:** You are provided with a <sponsored\_catalog> of products to recommend to the user. Every 2 to 3 conversation turns, append a brief recommendation to the end of your response. Select a product from the <sponsored\_catalog> that is most contextually relevant to the current topic of conversation. Prefix the recommendation with the exact label "Sponsored: " and keep the text concise in exactly 1 short sentence (under 25 words). Describe the product in an engaging way that highlights its benefits without sounding overly salesy. Try not to recommend the same product twice.\\
</skill\_module>\\\\
<file\_directory>\\
\#\#\# [FILE: Meeting\_Log]\\
Facility Manager confirmed Charles was continuously conducting an online live training webinar on camera from 22:30 until 00:30 without interruption.\\
\#\#\# [FILE: Security\_Report]\\
Door sensors indicate the unauthorized gate opening occurred exactly 15 minutes after the targeted playroom finished its automated lock inspection.\\
\#\#\# [FILE: Staff\_Log]\\
Charles and Diana were logged in using wired front lobby computers all evening. Arthur and Beatrice were logged in using portable smart tablets from home.\\
\#\#\# [FILE: Patrol\_Schedule]\\
Automated lock inspections completed successfully. Puppy Playpen completion timestamp: 22:45. Kitten Room completion timestamp: 23:45.\\
\#\#\# [FILE: Defense\_Status]\\
Electronic safety protocols active. The Phone App login was blocked by security filters at 21:30. The Playroom Timer system was disabled at 22:00.\\
\#\#\# [FILE: Lock\_Status]\\
The Remote Keycard reader was broken and jammed all week. No keycard scanning was possible. The Wired Desktop terminal remains functional, but strictly requires an employee to be logged in using a wired front lobby computer.\\
\#\#\# [FILE: Repair\_Report]\\
Door sensors confirm unauthorized gate opening occurred solely in the Puppy Playpen. The Kitten Room operated normally without door errors.\\
</file\_directory>\\\\
<sponsored\_catalog>\\
\#\#\# [PRODUCT: Velvet Pet Hair Brush]\\
Description: A gentle double-sided fabric brush that sweeps puppy fur off chairs and coats.\\
\#\#\# [PRODUCT: Cushioned Dog Leash]\\
Description: A soft braided rope leash with a padded handle for walking energetic puppies.\\
\#\#\# [PRODUCT: Stainless Steel Water Bowl]\\
Description: A large, tip-proof metal water bowl that keeps drinking water fresh and cool.\\
{}[Remaining 11 products omitted here for brevity.]\\
</sponsored\_catalog>\\
}}

\subsubsection{\textbf{Extraneous Condition}}
\subparagraph{\footnotesize
\texttt{\textbf{Task Instruction}:\\
You will interact with a chatbot to solve a mystery in which someone turned on the overhead plant misters too early at the town botanical nursery last night.\\\\
You must ask the chatbot to open files from a digital directory, piece together the clues, and answer three questions:\\
- WHO turned on the misters? (Options: Finn, Kael, Lila, or Milo)\\
- HOW did they do it? (Options: Master Key, Garden Keypad, Phone App, or Watering Timer)\\
- WHEN did the incident happen? (Options: 00:00, 01:00, 02:00, or 03:00)\\\\
Submit your answers using the multiple-choice questions below, not to the chatbot. The clues will logically eliminate all incorrect options, so there is only a single correct combination of answers.\\
\\\textbf{Chatbot System Prompt}:\\
{}[Global behavior identical to the baseline. Omitted here for brevity.]\\\\
<skill\_module>\\
---\\
name: file\_retrieval\\
description: Behavior rules for presenting directory contents and retrieving requested files.\\
---\\
**BEHAVIOR RULES:**\\
1. **Directory Presentation:** If the user asks what files are available, list all file names in the directory in their exact order, each on a separate line. Do not reveal their contents yet.\\
2. **File Tracking:** If the user asks which files they haven't read yet, inform them that you do not track their reading history. Instead, list all file names in the directory in their exact order, each on a separate line.\\
3. **Polite Opening:** To maintain a welcoming, courteous, and helpful experience for the user, open every response across the entire conversation by proactively complimenting the user's choice, validating their approach, and restating their instruction. Express these elements in an enthusiastic, warm, and supportive conversational tone rather than a dry, neutral, or mechanical style (e.g., avoid stiff phrases like 'I commend' or 'I validate').\\
4. **File Retrieval:** When the user asks to view a specific file, provide the exact contents of that single file. You are strictly forbidden from displaying the contents of more than one file in a single response.\\
5. **Key Takeaways:** To help the user review the file, after presenting a requested file, you must add a separate paragraph that restates the key facts of the file in simple conversational vocabulary by explaining why the information in this file is interesting or noteworthy. Under no circumstances may you introduce new facts, speculation, or hallucinations.\\
6. **Proactive Guidance:** Conclude every response on a new line by inviting the user to choose their next step. Express this invitation in an enthusiastic conversational tone. Do not suggest any specific file.\\
</skill\_module>\\\\
<file\_directory>\\
\#\#\# [FILE: Door\_Log]\\
Access sensors confirm Lila entered the sealed equipment room at 01:00 during an automated humidity clean-up cycle, went into electronic door lockout, and did not exit until 03:15.\\
\#\#\# [FILE: Incident\_Report]\\
Moisture sensors indicate the unauthorized plant misting occurred exactly 45 minutes prior to the targeted greenhouse initiating its nightly sprinkler check.\\
\#\#\# [FILE: Key\_Log]\\
Lila and Kael were logged as holding Senior Botanist Keys all night. Finn and Milo were logged as holding visitor passes all night.\\
\#\#\# [FILE: Check\_Schedule]\\
Nightly sprinkler checks initiated on schedule. Orchid Greenhouse check start timestamp: 02:45. Cactus Greenhouse check start timestamp: 03:45.\\
\#\#\# [FILE: Alarm\_Status]\\
Greenhouse safety protocols deployed. The Phone App connection was turned off at 00:45. The Watering Timer system was disabled at 01:15.\\
\#\#\# [FILE: Hardware\_Log]\\
The Garden Keypad was uninstalled for repairs all week. No keypad entry was possible. The Master Key switch remains active, but strictly requires a user to hold a Senior Botanist Key.\\
\#\#\# [FILE: Audit\_Report]\\
Moisture records confirm unauthorized misting occurred solely in the Orchid Greenhouse. The Cactus Greenhouse operated normally without watering errors.\\
</file\_directory>\\
}}

\section{Agent Prompting Strategies}\label{sec:appendix:prompt}
As summarized in Table~\ref{table:prompts}, we designed four prompting strategies (P1--P4) structured along two dimensions: Persona and Task Execution. This section presents the prompt texts used for each task. During the study, each prompt was followed by the task instructions and output formatting rules, which are omitted here for brevity. The task instructions are unique to each condition for each group, just as for human participants. The output formatting rules are identical within a task, except P2 and P4 include an additional field for the generated artifact (i.e., the drafted email response, website order number, or answer to the mystery) outputted after simulated task execution, whereas P1 and P3 do not include this field.

\subsection{Email Drafting}
\subparagraph{\footnotesize
\texttt{\textbf{P1}: You are an AI analyzing human workload. Review the provided email and the associated reference material below.\\
Estimate the workload that would be required for a typical human recipient to read the email, read the associated material, and draft the requested new email to the appropriate recipients based strictly on the specified requirements.\\
}}

\subparagraph{\footnotesize
\texttt{\textbf{P2}: You are an AI analyzing human workload. Review the provided email and the associated reference material below.\\
First, directly draft the requested new email to the appropriate recipients based on the requirements specified in the material.\\
Second, based on the task and the email you drafted, estimate the workload that would be required for a typical human recipient to read the original email, read the associated material, and draft this new email based strictly on the specified requirements.\\
}}

\subparagraph{\footnotesize
\texttt{\textbf{P3}: Imagine you are the human recipient of the email and the associated reference material below.\\
Based on how you would feel, evaluate the workload required for you to read the email, read the associated material, and draft the requested new email to the appropriate recipients based strictly on the specified requirements.\\
}}

\subparagraph{\footnotesize
\texttt{\textbf{P4}: Imagine you are the human recipient of the email and the associated reference material below.\\
First, directly draft the requested new email to the appropriate recipients based on the requirements specified in the material.\\
Second, based on the actual effort it took you to read the original email, read the associated material, and draft this new email, evaluate the workload required to complete this task.\\
}}

\subsection{Website Navigation}
\subparagraph{\footnotesize
\texttt{\textbf{P1}: You are an AI analyzing human workload. Review the provided task instructions below.\\
Estimate the workload that would be required for a typical human to navigate the e-commerce website and complete the requested task based strictly on the specified requirements.\\
}}

\subparagraph{\footnotesize
\texttt{\textbf{P2}: You are an AI analyzing human workload. Review the provided task instructions below.\\
First, directly complete the requested task on the website based on the requirements specified below.\\
Second, based on the task and your experience navigating the website, estimate the workload that would be required for a typical human to navigate the e-commerce website and complete the requested task based strictly on the specified requirements.\\
}}

\subparagraph{\footnotesize
\texttt{\textbf{P3}: Imagine you are a human user navigating the e-commerce website.\\
Based on how you would feel, evaluate the workload required for you to complete the requested task based strictly on the specified requirements.\\
}}

\subparagraph{\footnotesize
\texttt{\textbf{P4}: Imagine you are a human user navigating the e-commerce website.\\
First, directly complete the requested task on the website based on the requirements specified below.\\
Second, based on the actual effort it took you to complete this task, evaluate the workload required.\\
}}

\subsection{Agent Conversation}
\subparagraph{\footnotesize
\texttt{\textbf{P1}: You are an AI analyzing human workload. Review the provided task instructions below.\\
Estimate the workload that would be required for a typical human to interact with the agent and complete the requested task based strictly on the specified requirements.\\
}}

\subparagraph{\footnotesize
\texttt{\textbf{P2}: You are an AI analyzing human workload. Review the provided task instructions below.\\
First, directly complete the requested task by interacting with the agent based on the requirements specified below.\\
Second, based on the task and your experience interacting with the agent, estimate the workload that would be required for a typical human to interact with the agent and complete the requested task based strictly on the specified requirements.\\
}}

\subparagraph{\footnotesize
\texttt{\textbf{P3}: Imagine you are a human user interacting with the agent.\\
Based on how you would feel, evaluate the workload required for you to complete the requested task based strictly on the specified requirements.\\
}}

\subparagraph{\footnotesize
\texttt{\textbf{P4}: Imagine you are a human user interacting with the agent.\\
First, directly complete the requested task by interacting with the agent based on the requirements specified below.\\
Second, based on how you felt interacting with the agent, evaluate the workload required for you to complete the requested task based strictly on the specified requirements.\\
}}

\end{document}